\documentclass[11pt,epsf,epsfig,psfrag]{article}
\usepackage{hyperref}
\usepackage[bf]{caption}
\usepackage{epsfig}
\usepackage{psfrag}
\usepackage{color}
\usepackage{amsmath}
\usepackage[mathscr]{euscript}
\usepackage[toc,page]{appendix}

\usepackage{titlesec}
\titleformat{\paragraph}
{\normalfont\normalsize\bfseries}{\theparagraph}{1em}{}
\titlespacing*{\paragraph}
{0pt}{3.25ex plus 1ex minus .2ex}{1.5ex plus .2ex}

\usepackage{float}
\usepackage{graphicx}
\usepackage{epstopdf}
\usepackage{subfig}
\def\ga{\mathrel{\raise.3ex\hbox{$>$\kern-.75em\lower1ex\hbox{$\sim$}}}}
\def\la{\mathrel{\raise.3ex\hbox{$<$\kern-.75em\lower1ex\hbox{$\sim$}}}}

\def\I_M{{I_{\scriptscriptstyle M\times M}}}

\def\lsim{\mathrel{\rlap{\lower4pt\hbox{\hskip1pt$\sim$}}
    \raise1pt\hbox{$<$}}}                
\def\gsim{\mathrel{\rlap{\lower4pt\hbox{\hskip1pt$\sim$}}
    \raise1pt\hbox{$>$}}}   

\begin{document}

\thispagestyle{empty}

\vskip 2cm

\begin{center}{\Large \bf Thermodynamic geometry of one-dimensional spin one lattice models}
\end{center}

\vskip .2cm

\vskip 1.2cm

\centerline{ \bf Anurag Sahay, Riekshika Sanwari\footnote{anuragsahay@nitp.ac.in, riekshika.ph19@nitp.ac.in  } 
}

\vskip 4mm\centerline{ \it  Department of Physics, National Institute of Technology, Patna 800005,  India}

\vskip 1.2cm
\vskip 1.2cm
\centerline{\bf Abstract}
\noindent

State space geometry is obtained for the one dimensional Blume Emery Griffiths model and the associated scalar curvature(s) investigated for various parameter regimes, including the Blume-Capel limit and the Griffiths model limit. For the one-dimensional case two complementary geometries with their associated curvatures $R_m$ and $R_q$ are found which are related to the fluctuations in the two order parameters, namely the magnetic moment and the quadrupole moment. An excellent agreement is obtained in significant regions of the parameter space between the two curvatures and the two corresponding correlation lengths $\xi_1$ and $\xi_2$. The three dimensional scalar curvature $R_g$ is also found to efficiently encode interactions. The scaling function for the free energy near critical points and the tricritical point is obtained by making use of Ruppeiner's conjecture relating the inverse of the singular free energy to the thermodynamic scalar curvature. 
\newpage
\setcounter{footnote}{0}
\noindent

\section{Introduction}
\label{intro}

Thermodynamic geometry (TG), pioneered by Ruppeiner and other workers, has been extensively used to probe a wide range of systems including fluids, magnetic systems and several black hole solutions. Employing a metric based on second moments of thermal fluctuations TG renders a Riemannian geometric structure to the thermodynamic state space of the system, \cite{rupprev,rupporiginal}. The thermodynamic metric quantifies in a co-ordinate independent way the classical distinguishability of thermodynamic states, so that the easier it is for two states to fluctuate into each other the shorter the separation between them and vice versa. Not surprisingly then, near the critical point where fluctuations diverge the separation between thermodynamic states shrinks and the associated scalar curvature diverges.

Beyond this association, TG forges a surprising and a remarkable connection between its geometric invariants which are calculated solely from thermodynamics and the statistical mechanical description of the system. Thus, in an early insight Ruppeiner conjectured that in the vicinity of the critical point the state space scalar curvature $R$ is $equal$ to the correlation volume $\xi^d$ upto a constant of order unity. In the asymptotically critical region this conjecture could be further refined to an equality of the scalar curvature with the inverse of the critical free energy $\psi_s$ upto a universal constant $\kappa$ which depends only on the universal critical exponents. This geometry-energy equation in the Riemannian state space is a centre-piece of TG and is reminiscent of the Einstein's equation which relates space-time curvature to matter energy distribution. It will therefore be apt to refer to the conjectured equality
\begin{equation}
R=\kappa\,\frac{1}{\psi_s}  
\label{rpsi}
\end{equation}
 with $\kappa$ a universal constant, as ``the Ruppeiner equation". The correspondence of $R$ with the correlation length then follows, at least in the near critical region, from the well known equality of the correlation volume with the inverse of singular free energy density. The Ruppeiner equation could be profitably used to calculate in a straightforward manner the scaling function of the singular free energy, thus providing an alternative to the more challenging calculation based on renormalization group analysis, \cite{ruppcrit,ruppspin}.

    It turns out, the Ruppeiner equation can be thought of as the stronger form of the conjecture relating geometry to thermodynamics, one which is exact $at$ the critical point. The conjecture relating curvature to correlation length as
 
 \begin{equation}
|R|\sim\xi^d
\label{rcorr}
\end{equation}   
    
     (where $d$ is spatial dimension) on the other hand is found to be valid for parameter values much beyond the critical point. We shall refer to it as the weak end of Ruppeiner's conjecture. It has been shown in this work as also earlier elsewhere that the connection between curvature and correlation length extends even to situations where there is no criticality. The weak conjecture has found much use in recent times in calculating the phase coexistence curves and the Widom lines for fluids and other systems, \cite{sahay1,may,sarkar1,ruppmay}. Remarkably, the scalar curvature packs even more information about the underlying statistics of interactions via its signature. Thus, it is commonly believed that a positive sign (in the convention used in \cite{rupprev}) of $R$ is indicative of statistically repulsive interactions while a positive sign is suggestive of statistically attractive interactions, \cite{ruppmay,mirza,rupptrip,rupprep}. For example, the scalar curvature has always been seen to diverge to negative infinity at criticality. However, the issue of signature of $R$ is nuanced as suggested recently in \cite{ruppbinary} and it still awaits a more fundamental resolution.  In addition, scalar curvature in three or higher dimensional parameter space has been investigated for only a handful of thermodynamic systems \cite{ruppbinary,rupphigher}. It is of interest to pursue the geometry of higher dimensional parameter spaces since apart from the scalar curvature the sectional curvatures could also signify underlying physics. More generally $R$ is to be understood as a qualified measure of interactions. While its behaviour is well understood broadly, a finer understanding of the geometric curvature is still an ongoing work. In light of this it is important to record the behaviour of $R$ for a range of parameter values in a given system, something this work attempts to do. 
    
 Exactly solved models can serve as important testing grounds to verify the conjectures of TG and also to further explore the features of $R$ inasmuch as they offer an analytical control over the partition function and also possibly the correlation length. One of the most important such models used to successfully verify TG is the one dimensional Ising model which has a (pseudo)critical point at zero temperature in zero magnetic field. The scalar curvature earlier worked out numerically in \cite{ruppmag} was later found to be a surprisingly simple expression, \cite{mrugala}. For zero field, $R$ tends to exactly twice $\xi$ as it nears the zero temperature pseudocritical point. Furthermore, for the non critical case in the presence of a magnetic field the curvature $R$ forms an umbrella over the correlation length and converges to $2$ as $\xi$ decays to zero at low temperatures (see figs.(\ref{ising1}) and (\ref{ising2})). Some other cases where TG has been applied to exactly solved models are the Ising model on a Bethe lattice, \cite{dolan1}, Ising model on planar random graphs, \cite{dolan2}, the spherical model, \cite{dolan3} and the one dimensional Potts model, \cite{dolan4}.

Ising model with lattice spins $S_i=\pm 1$ naturally generalizes to the spin one model with lattice spins $S_i=0,\pm 1$. What might seem like an innocuous addition of a degree of freedom at lattice sites results in a rich and varied phase structure of the spin one model. This is because, apart from the spin-spin quadratic coupling and the coupling of individual spins to the magnetic field $H$, the spin one model admits the possibility of a non trivial biquadratic coupling of quadrupoles $S_i^2$ to each other and their coupling to an ordering field independent of the magnetic field. In effect, the thermodynamics of a spin one model is governed by two order parameters, the spin $\langle S_i\rangle$ and the quadrupole moment $\langle S_i^2\rangle$. Both the order parameters, while kinematically coupled, are separate stochastic variables and their interplay leads to a rich phase structure with coexistence surfaces bordered by lines of critical points and first order points which meet in one or more tricritical points.
Naturally, therefore, spin one lattice models have been extensively used to model the behaviour of interacting systems with two types of ordering processes. One of the most popular spin one modes, the Blume-Emery-Griffiths (BEG) model originally formulated to study the phase behaviour  $\mbox{He}^3-\mbox{He}^4$ mixtures, has been widely used to model diverse phenomena. As a model for the phase behaviour in Helium mixture it successfully captures the phenomena of superfluid ordering as well as phase separation depending on the relative concentration of the  $\mbox{He}^3$ impurity, \cite{beg}. In addition the BEG model has been used in the context of simple fluids to model condensation and solidification, \cite{sivar}, in binary alloys to model ferromagnetism and phase separation, \cite{rys}, and in microemulsions, \cite{microem}, to name a few. In the limit of zero biquadratic coupling, known as the Blume-Capel model, \cite{blume,capel}, it has been used to model
phase behaviour in magnetic systems wherein depending on the strength of crystal field splitting the transition between a paramagnet and a magnetically ordered state changes from first order to second order.  

In this work we initiate the study of TG for classical higher spin lattice models beginning with an investigation of the  one dimensional spin one models, namely the Blume Emery Griffiths model and its limiting cases of the Blume Capel model and the Griffiths model,\cite{griff}, where in the latter case the quadratic spin coupling is set to zero. In addition to the advantage of being exactly solved, the one dimensional spin one models retain much of the rich phase behaviour of their higher dimensional counterparts. Thus, these models continue to display a locus of pseudocritical points, a pseudotricritical point and zero temperature phase coexistence, much of which is amenable to a geometric treatment. Besides, the parameter space of BEG models is three dimensional which provides an avenue to explore higher dimensional scalar curvature and various sectional curvatures. In addition, the fact that there are two order parameters in the model gives rise to the possibility of two correlation lengths for some parameter values and it would be worthwhile exploring if geometry encodes different correlation lengths. In this work we hope to make good use of the given opportunity.

This paper is organised as follows. In section \ref{spin one} we first discuss the Hamiltonian of the Blume Emery Griffiths model and follow it up with a discussion of its mean field phase structure. We then review the one dimensional BEG model in the framework of the transfer matrix solution of its partition function and subsequently review its phase structure. In  section \ref{hyper} we start with a tutorial introduction to the geometrical representation of thermodynamic constraints and then argue our case for two hypersurface geometries we believe are most relevant to our model. Finally, in section \ref{onedgeo} we present our main results for the one dimensional BEG model. In subsection \ref{scaling} we obtain the singular part of free energy for the case of positive spin-spin coupling and using TG we work out on the spin scaling function for the one dimensional BEG model. In subsection \ref{jzero} we obtain the geometry of the Griffiths model, with a special emphasis on the three dimensional scalar curvature. Finally in section \ref{conclu} we summarize our main finding and scope of work and point to some future directions.

\section{The spin one model}
\label{spin one}

The BEG model has the most general reflection symmetric Hamiltonian for a classical spin one model with nearest neighbour interactions. The Hamiltonian for the BEG model is written as, \cite{beg}
\begin{equation}
 \mathcal{ H}_{beg} = -J\, \sum_{<ij>} \, S_i\, S_{j}-H\, \sum_{i} \, S_i  - K\,\sum_{<ij>}\, S_i^{\,2}\, S_{j}^{\,2}+ D\,\sum_{i} S_i^{\,2}
\label{BEG}
\end{equation}
The lattice spin variable $S_i $ is Ising like and can take up values $+1,-1$ and $0$. In addition to the bilinear coupling terms and a magnetic field that couples to the magnetic moment, the Hamiltonian contains a biquadratic coupling term of strength $K$ and a crystal field $D$ which couples to the quadrupole moment. The coupling strengths $J$ and $K$ are positive in the original BEG model, \cite{beg}. The $K=0$ limit is the Blume Capel (BC) model, \cite{blume,capel}. While the magnetic field term is not experimentally realizable in the original context of the BEG model which refers to a mixture of  $\mbox{He}^3-\mbox{He}^4$, it plays its usual role in the BC limit which refers to a magnetic system. The spin one models have two densities, namely the mean magnetization and the mean quadrupole moment,
\begin{equation}
 M=\langle S_i\rangle\,\,\,\,;\,\,\,\, Q=\langle S_i^2\rangle
\label{order}
\end{equation}
Since the model is translationally invariant, there is no spatial variation in the order parameters. In the lattice gas interpretation of the BEG model $M$ represents the superfluid order parameter while $Q$ is the concentration of $\mbox{He}^4$. In a magnetic system $x=1-Q$ would measure the concentration of non magnetic impurities. Similarly, apart from a small term, $D$ is the difference in chemical potentials of $\mbox{He}^3$ and $\mbox{He}^4$, namely $D \sim \mu_3-\mu_4$. For a positive $D$ larger concentrations of $\mbox{He}^3$ would be energetically preferred. In the context of a magnetic system, $D$ refers to a single ion anisotropy term  which splits the single-spin energy levels, with $S_i =0$ lower than the degenerate $S_i=\pm1$ levels. It can also be thought of as an external field coupled to the order parameter $Q$ (or $x$) analogous to $H$ which couples to $M$. Owing to the interplay of the two order parameters  via the interaction and field terms the phase structure of the BEG and related models is rich, with the presence of tricritical points, critical lines and a line of first order transitions, \cite{beg,blume,capel}. We now sketch the phase behaviour very briefly.

\begin{figure}[t!]
\centering
\includegraphics[width=3in,height=2.5in]{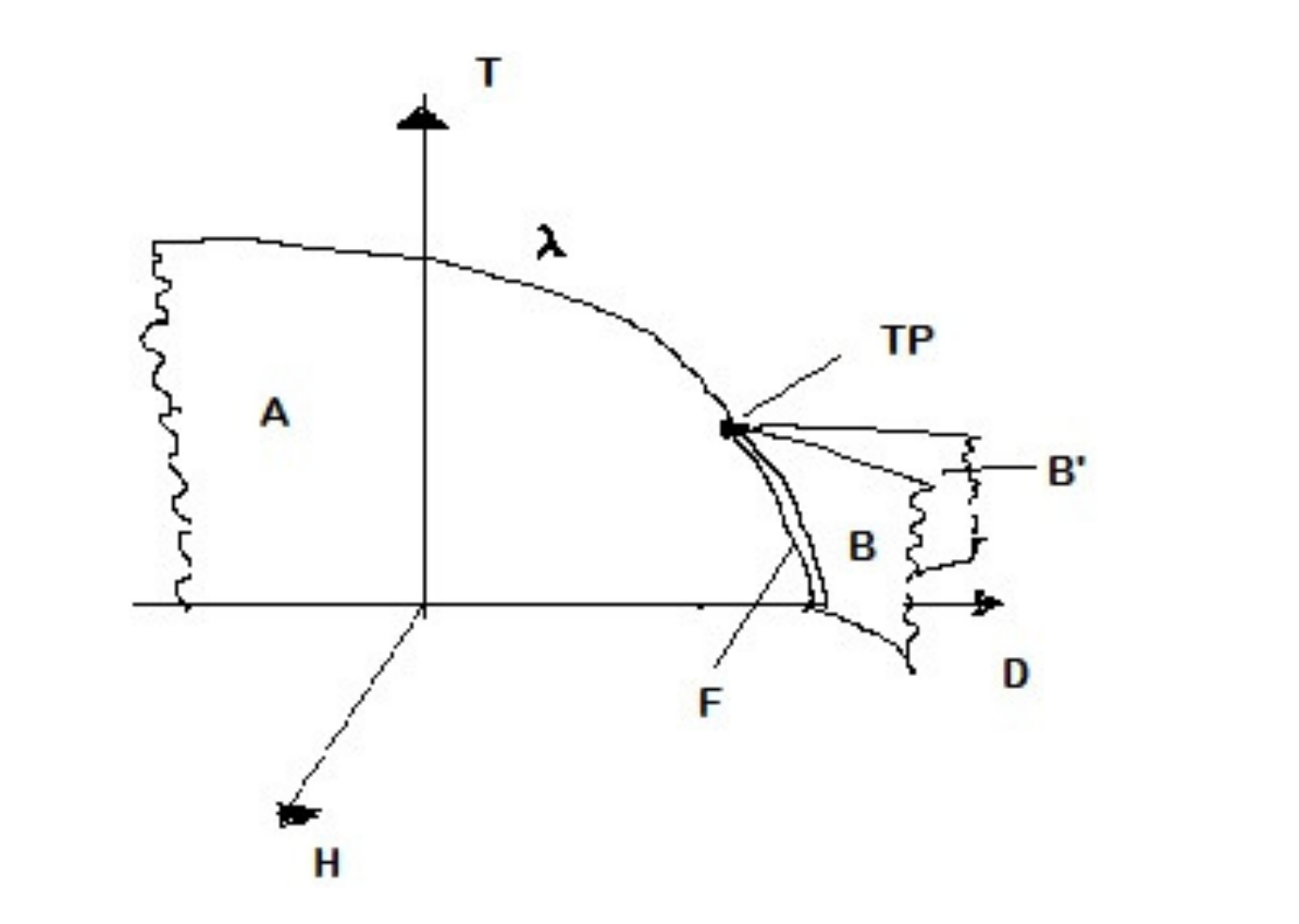}
\caption{Schematic diagram of the mean field phase structure of the spin one model in the $T-D-H$ plane. Here the ratio $K/J$ is small.}
\label{meanfieldbeg}
\end{figure}
 
Fig.(\ref{meanfieldbeg}) above shows a mean field picture of the phase diagram of the spin one model with the Hamiltonian given by eq.(\ref{BEG}) and with the parameter ratio $K/J$ small which relates to the physial context of the BEG model, \cite{beg,mukamel}. ${\bf A}$ is a coexistence surface in the $T-D$ plane where phases with positive and negative $M$ coexist. For $D\to-\infty$ the $S=0$ state is completely suppressed and the system can be mapped to a spin half Ising model. In the $\mbox{He}^3-\mbox{He}^4$ context it would mean the presence of only the $\mbox{He}^4$ state. In addition to the ${\bf A}$ surface there are symmetrically placed wing like coexistence surfaces ${\bf B}$ and ${\bf B'}$ that extend into the $D-H$ plane for $D>0$. On each of these surfaces two phases with different values of $Q$ coexist. The three coexistence surfaces each terminate on the high temperature side in lines of critical points, which intersect and terminate at the tricritical point {\bf TP}. The surfaces ${\bf A,B}$ and ${\bf B'}$ intersect in the $T-D$ plane along a line ${\bf F}$ of first order phase transitions which in turn terminates at the tricritical point. Three phases coexist on ${\bf F}$ so that it is a line of triple points.  The tricritical point is different from the critical point in that the critical exponents $\beta$ and $\delta$ are different from their respective mean field values. The phase diagrams change for higher values of the ratio $K/J$. Thus, for example, at $K/J=3.1$ there are two more tricritical points symmetrically positioned on the critical boundaries of the wings ${\bf B}$ and ${\bf B'}$, \cite{mukamel}. Furthermore, anti-ferromagnetic bilinear coupling or a repulsive biquadratic term which render the ratio $K/J<0$, diversifies and further enriches the phase behaviour in the spin one model, \cite{hoston,wang}. In the limit $J=0$ which we call as the Griffiths model, the BEG model exhibits phase coexistence as well as second order transition in the quadrupolar order parameter, \cite{griff}. In this limit the spin one model can be mapped onto the spin $\frac{1}{2}$ Ising model with a temperature dependent magnetic field.  We shall not be pursuing the geometry of antiferromagnetic spin coupling or a repulsive biquadratic coupling in this work. While we believe the geometry of these cases is very interesting, it is more nuanced due to the presence of staggered spin and quadrupolar orders. We shall return to these exciting cases in our future work.

It may be be noted that the most general nearest neighbour spin one Hamiltonian can be written by adding an asymmetric exchange term between the dipole and the quadrupole moments. This was found by Mukamel and Blume (MB) \cite{mukamel,krinsky},
\begin{equation}
 \mathcal{ H}_{mb} = \mathcal{H}_{beg}-\frac{L}{2}\sum_{<ij>}(S_i\,S_j^{\,2}+S_i^{\,2}\,S_j)
\label{s 1 ham}
\end{equation}
It can be seen that for $L>0$ neighbouring states with $S_i=1$ will be energetically most preferred while those with $S_i=-1$ will be least preferred thus breaking the up-down symmetry. Staggered quadrupolar order is observed in this case for a range of values of $L$, \cite{mukamel,krinsky}. For the same reason as above we set aside a geometric investigation of the MB model to a future work.

\subsection{The one dimensional spin one model.  }
\label{oned}

 The most general Hamiltonian of a spin one chain of N atoms with nearest neighbour interaction is
\begin{equation}
\mathcal{H}= - J\, \sum_i^N \, S_i S_{i+1} + K\,\sum_i^N S_i^2 S_{i+1}^2-  H\, \sum_i^N \, S_i-D\, \sum_i^N S_i^2 -\frac{L}{2}\sum_{i}^N(S_i\,S_{i+1}^{\,2}+S_{i+1}^{\,2}\,S_i)
\label{ham1d}
\end{equation}

Here the coupling constants and fields have their usual meaning as in the 
Hamiltonian, eq.(\ref{BEG}). The one dimensional ring of $N$ spins is exactly solvable in the large $N$ limit using the standard transfer matrix technique, \cite{krinsky}. The transfer matrix for the Hamiltonian in eq.(\ref{ham1d}) is three dimensional,
\begin{equation}
T=\left(
\begin{array}{ccc}
 e^{(-D +H  +J +K +L )\beta } & e^{(H-D) \beta /2} & e^{(-D -J +K) \beta } \\
 e^{(H-D)\beta /2 } & 1 & e^{-(D+H)\beta /2 } \\
 e^{(-D -J +K) \beta } & e^{-(D+H)\beta /2 } & e^{(-D -H +J +K -L) \beta } \\
\end{array}
\right)
\label{transfer}
\end{equation}

This gives rise to three eigenvalues in general, $\lambda_1>\lambda_2>\lambda_3$. In the limit of infinite $N$, i.e the thermodynamic limit, the logarithm of the largest eigenvalue $\lambda_1$ becomes the free energy per spin. The free energy (Massieu function) for the zero field BEG model, $L=H=0$ can be obtained in a closed form,
\begin{equation}
\psi=\mbox{ln}\left[\frac{1}{2} e^{-\beta  (D+J)} \left(e^{\beta  D+\beta  J}+e^{2 \beta  J+\beta  K}+e^{\beta  K}+\sqrt{W}\right)\right]
\label{psibegz}
\end{equation}
 where 
\begin{equation}
 W=\left(e^{\beta  D+\beta  J}+e^{2 \beta  J+\beta  K}+e^{\beta  K}\right)^2-4 \left(-2 e^{\beta  D+2 \beta  J}+e^{\beta  D+\beta  J+\beta  K}+e^{\beta  D+3 \beta  J+\beta  K}\right)\nonumber
\end{equation}
  while for the more general cases one can find it numerically by solving a cubic equation for the eigenvalues. In the following the BEG model ($L=0$) shall be our default model and we shall no more mention the Mukamel-Blume model ($L\neq 0$).

 While there is no finite temperature phase transition for the one dimensional case, its thermal behaviour richly responds to the interplay of various coupling strengths in the Hamiltonian. The phase diagram can be seen as a limiting case of the mean field phase structure as represented in fig.(\ref{meanfieldbeg}). In the limit of one dimension the critical points and the tricritical point will lie at lower and lower temperature until they become respectively pseudocritical points and the pseudotricritical point at $T=0$. In a sense the whole phase diagram of fig.(\ref{meanfieldbeg}) flattens out onto the $H-D$ plane, \cite{krinsky,mukamel}. At the pseudocritical point the eigenvalues $\lambda_1$ and $\lambda_2$ become asymptotically equal. At the pseudotricritical point all the three eigenvalues become asymptotically equal. 
 
 The correlation length can be obtained in a standard manner via the ratio of the largest and the next-to-largest eigenvalues. Interestingly, owing to the fact that there are two correlation functions corresponding to the spin-spin and the quadrupole-quadruple correlations, the spin one model admits the possibility of two separate correlation lengths. Indeed, this possibility is realized for the zero field BEG model with $H=L=0$ where due to increased symmetry of the transfer matrix there are separate correlation lengths for spin and quadrupole fluctuations, given respectively as,
\begin{eqnarray}
\xi_1^{-1}&=&\mbox{log}\frac{\lambda_1}{\lambda_2}\nonumber\\
\xi_2^{-1}&=&\mbox{log}\frac{\lambda_1}{\lambda_3}
\label{corr}
\end{eqnarray}
For non-zero $H$ field there is only one correlation length $\xi_1$ for correlations in both order parameters, \cite{krinsky}. It will be of interest to see how far the state space geometry encodes the microscopic statistical interactions of the chain. In particular one would like to find out if the state space scalar curvature is representative of the two correlation lengths $\xi_1 \mbox{and} \xi_2$ and as to how successful it is in encoding the pseudo critical and tricritical behaviours in the model. As reasoned out in the previous section we shall be exploring two two-dimensional geometries for the BEG model, namely the ones intrinsic to the $D$-surface and the $H$-surface. We shall also comment briefly on the full three dimensional geometry. Quite satisfyingly, we shall find that the two geometries faithfully represent the underlying correlations in the two order parameters.

\begin{figure}[!t]
\begin{minipage}[b]{0.5\linewidth}
\centering
\includegraphics[width=2.8in,height=2.3in]{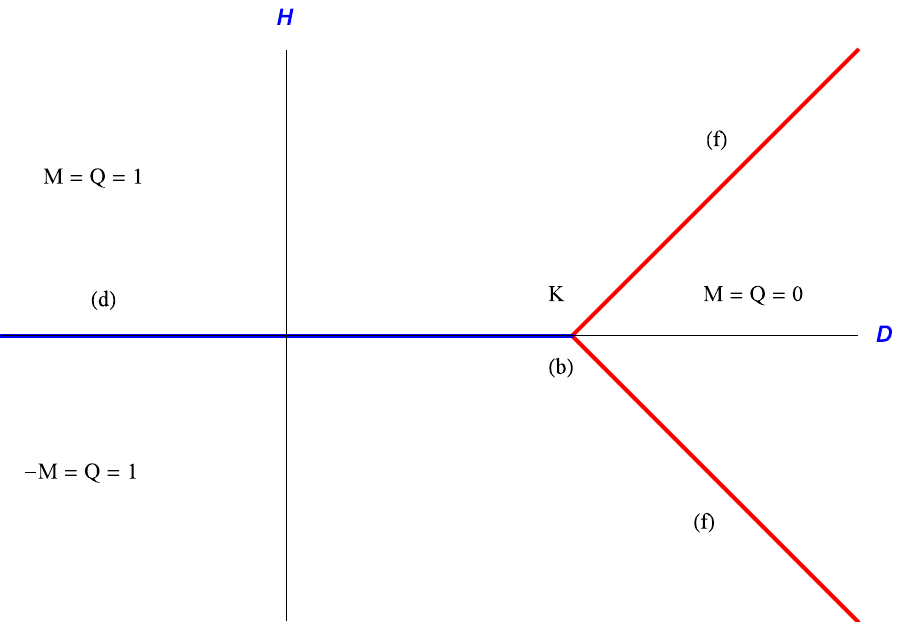}
\caption{\small{Zero temperature phase diagram for the spin one model in the $H-D$ plane with $L=0,J>0,J+K>0$. Two phases coexist on the $a$ line and the $f$ lines in the $T=0$ place and they are a locus of pseudocritical points on approaching from $T>0$. Point $c$ is a triple point at which three phases coexist in the $T=0$ plane. On approaching from $T>0$ it shows pseudotricritical behaviour. Adapted from \cite{krinsky}. }}
\label{krinA}
\end{minipage}
\hspace{0.6cm}
\begin{minipage}[b]{0.5\linewidth}
\centering
\includegraphics[width=2.8in,height=2.3in]{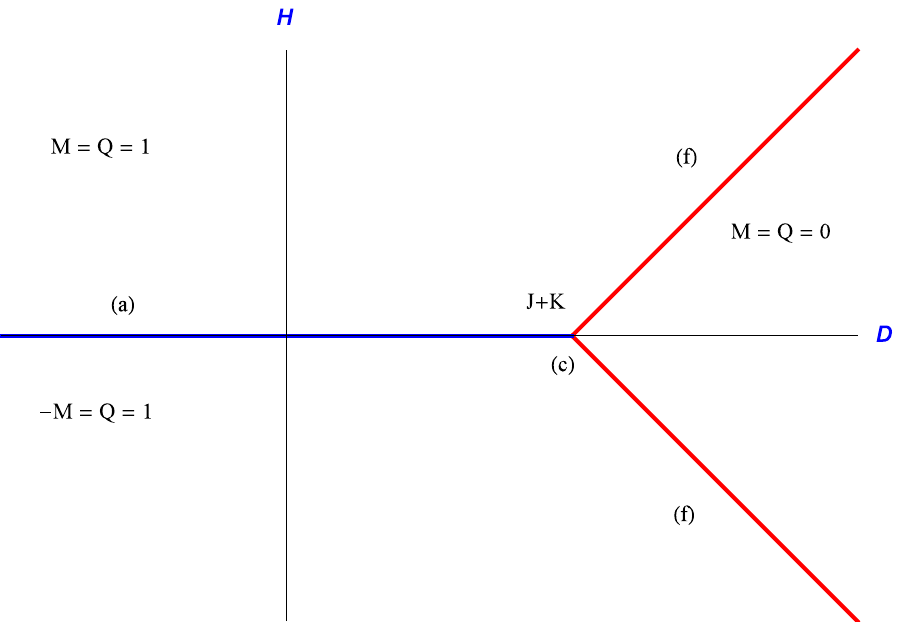}
\caption{\small{Zero temperature phase diagram of the spin one model in the in the $H-D$ plane with $L=J=0,K>0$. Infinite phases coexist on the line $d$ and no pseudocritical behaviour is seen here. Two phases coexist along the $f$ lines and pseudocritical behaviour is seen on approaching from $T>0$. The point $b$ is a triple point but not a pseudotricritical point. Adapted from \cite{krinsky}.}}
\label{krinB}
\end{minipage}
\end{figure}

Let us first briefly discuss the phase structure for the one dimensional BEG model. Our review closely follows the exposition in \cite{krinsky} and even the figures \ref{krinA} and \ref{krinB} are exactly similar to those in \cite{krinsky} down to the figure labels. In any case the diagrams are easily obtainable from the transfer matrix.
As mentioned earlier, in this report we shall restrict ourselves to the case of non-negative values of spin exchange and biquadratic exchange, namely $K,J\geq 0$. Fig. (\ref{krinA}) shows the phase diagram of the full one-dimensional BEG model at $T=0$. The phase diagram remains the same in the Blume-Capel limit $K=0$. Two phases with $Q=1$ and $M=\pm 1$ coexist on the $a$ line that is positioned at zero magnetic field. Similarly, two phases $Q,M=0$ and $Q=1,M=\pm 1$ coexist on the two symmetrically placed $f$ lines. There is a discontinuous phase change on crossing these lines at zero temperature whereas on approaching these lines from $T>0$ one sees pseudocritical behaviour. The intersection of the $a$ line with the $f$ lines is the triple point $c$ at which three phases coexist. On approaching the $c$ line from $T>0$ pseudotricritical behaviour is seen. Fig. (\ref{krinB}) is the zero temperature phase diagram of the BEG model in the limit of zero quadratic coupling, $J=0$. On the $d$ line an infinite number of phases coexist in the limit of infinite $N$. This is because, for $D<K$ on the $d$ line the $S=0$ spin is energetically ruled out while the up and down spins are equally likely in the absence of a discriminating quadratic coupling. In other words, the model behaves like a collection of independent Ising spins, which also rules out any pseudocritical behaviour in approaching the line from $T>0$. On the other hand, the $f$ line continues to show two-phase coexistence just as for the case $J>0$ as well as pseudocritical behaviour on approaching from $T>0$. The point $b$ is a triple point but not a pseudotricritical point. The phase diagram of the spin one model with the spin-spin coupling strength $J$ set to zero is presented in fig.(\ref{krinB}) which therefore represents the zero temperature phase diagram of the one-dimensional Grifiths model. 

 Understandably, the zero temperature phase diagram in fig.(\ref{krinA}) and fig.(\ref{krinB}) is straightforward, with straight line phase boundaries, because it is influenced only by energy considerations in the absence of any thermal or quantum fluctuation. The region to the right of  
the $f$ lines in both figures is the one where the $D$ `field' in the Hamiltonian, eq.(\ref{ham1d})  is greater than the combined strength of the attractive couplings $J$ and $K$, so that the energetically preferred state is the one with all spins set to zero.

\section{The geometry of thermodynamic constraints: a tutorial introduction }
\label{hyper}

In this section we explain our method of carving out the relevant state space geometries when restrictions are applied on some thermodynamic fluctuations in any system. The method of constructing state space hypersurfaces corresponding to thermodynamic constraints was first discussed in the context of black hole thermodynamics in \cite{sahay restricted}.
 While our reference system shall be the BEG model the general method is applicable to any system with three or more thermodynamic variables. 
 
 We first set up the equilibrium thermodynamic relations beginning with the partition function. While basic and well known in the context of magnetic systems the ensuing discussion is of fundamental value to the subsequent construction of the model's thermodynamic geometry. 
The canonical partition function per site of the BEG model is a weighted sum over all possible spin configurations,
\begin{equation}
Q=\frac{1}{N}\sum_{\left\lbrace S_i \right\rbrace}\mbox{exp}^{-\beta\,( -J\, {\sum_{<ij>}} \, S_i\, S_{j}- K\,\sum_{<ij>}\, S_i^{\,2}\, S_{j}^{\,2}-H\, \sum_{i} \, S_i + D\,\sum_{i} S_i^{\,2} )}
\label{partfunct}
\end{equation}
The Massieu function per spin $\psi$, expressed as the log of the partition function, can be taken as the starting point for thermodynamics and also for the geometry. The state space Riemannian metric is conveniently defined as the second deriative of the Massieu function with respect to its entropic intensive variables. Note that throughout this text we shall interchangeably use the terms ``free energy'' and Massieu function so that by both terms we mean the log of the partition function. Indeed, ``free entropy'' would have been more apt here and we hope to be forgiven for this abuse of terminology.

It shall be profitable in the following to think of the parameter $D$ as a tunable external field, analogous to the magnetic field $H$. At the same time the self interaction parameters $J,K$ could be thought of as fixed for a given system. The macroscopic energy per site $E$ is obtained as the derivative of the Massieu function
\begin{eqnarray}
-\left.\frac{\partial{\psi}}{\partial\beta}\right|_{H,D}&=&E=\frac{1}{N}\langle{\mathcal{H}}_{beg}\rangle\nonumber\\
&=& \frac{1}{N}\langle -J\, \sum_{<ij>} \, S_i\, S_{j} - K\,\sum_{<ij>}\, S_i^{\,2}\, S_{j}^{\,2}\,\rangle-\frac{1}{N}\langle H\sum_i S_i\rangle +\frac{1}{N}\langle D\,\sum_i S_i^2\rangle\nonumber\\
&=& \langle\mathcal{F}_1\rangle+\langle\mathcal{F}_2\rangle+\langle\mathcal{F}_3\rangle\nonumber\\
&=& U -H\,M+D\,Q
\label{E def}
\end{eqnarray}
where
\begin{eqnarray}
U&=&\langle -J\, \sum_{<ij>} \, S_i\, S_{j} - K\,\sum_{<ij>}\, S_i^{\,2}\, S_{j}^{\,2}\,\rangle/N\nonumber\\,
M&=&\langle\sum_i S_i\rangle/N\,\,\,\, \mbox{and}\nonumber\\
Q&=&\langle\sum_i S_i^2\rangle/N
\label{UMQ}
\end{eqnarray}
with $N$ being the total number of spins. Eq.(\ref{E def}) above shows a natural breakup of the Hamiltonian into three stochastic variables $\mathcal{F}_1,\mathcal{F}_2\, \mbox{and}\,\mathcal{F}_3$ whose mean values are related to the thermodynamic quantities, $U,M$ and $Q$. This can be compared to the Hamiltonian of the spin half Ising model which breaks up into two stochastic variables, \cite{mrugala}.  The quantity $U$ in eq.(\ref{UMQ}) above could be thought of as the``\,internal" energy per site of the system comprising self interaction terms in the Hamiltonian while $E$ is the ``\,total" energy per site which includes the interaction energy between the spin system and the sources of external fields \cite{callen}. The three correlated but independent fluctuations render the geometry of the spin one model three dimensional.

The entropy per site $S$ is obtainable from the partition function via the specific Massieu function
\begin{eqnarray}
\psi&=& S-\beta\,E\nonumber\\
&=& S-\beta\,U-\nu\,M-\mu\,Q
\label{psiE}
\end{eqnarray}
where in the second equality we have used the entropic intensive variables $\nu=-\beta H$ and $\mu=\beta D$. 

 Using the partial derivatives of $\psi$ with respect to $\beta$, $H$ and $D$ the differential of $\psi$ is obtained as
\begin{equation}
d\psi(\beta,H,D)=-E\,d\beta+\beta\, M\,dH-\beta\,Q\,dD
\label{dpsiE}
\end{equation}
Eqs.(\ref{dpsiE}) and  the first equality in (\ref{psiE}) lead to the first law for variations in the total energy $E$,
\begin{equation}
dS=\beta\,dE + \beta\,M\,dH-\beta\,Q\,dD
\label{dE}
\end{equation}
On the other hand, considered as a function of the entropic intensive variables $\beta$, $\nu$ and $\mu$ the differential of $\psi$ becomes 
\begin{equation}
d\psi(\beta,\nu,\mu)=-U\,d\beta\,-M\,d\nu-\,Q\,d\mu
\label{dpsiU}
\end{equation}
and the first law for variations in the internal energy $U$ is obtained as
\begin{equation}
dS=\beta\,\,dU-\beta\,H\,dM+\beta\,D\,dQ
\label{dU}
\end{equation}
In either case we of course find that $d^2 S=- d^2\psi$. We shall always take eq.(\ref{dpsiU}) and eq.(\ref{dU}) as our starting point for thermodynamic geometry.

The fundamental thermodynamic equation of the spin one system in the entropy representation is formally the function 
\begin{equation}
S=S(U,M,Q)
\label{fundaS}
\end{equation}
 from where all the thermodynamic quantities can be obtained by suitable differentiation, in particular, the equations of state, 
 \begin{eqnarray}
 T&=& T(U,M,Q)\nonumber\\
 H&=& H(U,M,Q)\nonumber\\
 D&=&D(U,M,Q).
 \label{eos}
 \end{eqnarray}
 Since the number $N$ of latttice sites is held constant it has already been absorbed in the definition of $M,Q,U$ and $S$ as per site thermodynamic quantities. In a real magnetic system the volume $V$ could be an additional parameter, with the background lattice serving to store energy. However, no such consideration is made for the spin one lattice which is a model magnetic system.

 A spin one system in equilibrium with fixed $T,H$ and $D$ can most easily be envisaged by considering it to be a small subsystem of a much larger spin system which acts as a $reservoir$.  This fixes the equilibrium densities of the extensive quantities $U$, $M$ and $Q$ of the subsystem to be the same as the reservoir values so that the global entropy of system plus reservoir is maximized. However, at any instant, the respective fluctuating random variables $\mathcal{F}_1, \mathcal{F}_2$ and $\mathcal{F}_3$ in the subsystem Hamitonian could take up values away from the global mean. Thus, in its three dimensional thermodynamic state space $\Gamma$ with axes $U,M,$ and $Q$ the subsystem would spontaneously fluctuate from its reservoir determined equilibrium point. It is important to remember here that these are equilibrium fluctuations so that even as the subsystem fluctuates away from the global average it always satisfies the fundamental thermodynamic equation, eq.(\ref{fundaS}).
 
  If all the fluctuations are unconstrained then the Riemannian metric of fluctuations is three dimensional. We take the Massieu function as our starting point and differentiate it twice with respect to the entropic intensive variables $\beta,\nu,\mu$ to generate the metric, which we call the $grand$ metric $g$. In our geometric notation we have  $x^1=\beta,x^2=\nu,x^3=\mu$, and for the extensive quantities $X_1=U,X_2=M,X_3=Q$. Then the grand metric is the covariance matrix for fluctuations in extensive variables,
\begin{equation}
g_{\,ij}=\frac{\partial^2\psi}{\partial x^i\,\partial x^j}=\langle\Delta X_i\,\Delta X_j\rangle
\label{fullmetric}
\end{equation}

A complimentary way of looking at fluctuations is via the three dimensional parameter space $\mathcal{M}$ with axes $T,H$ and $D$ or, equivalently, $\beta,\nu,\mu$. Thus, when the subsystem undergoes a spontaneous fluctuation its $intrinsic$ intensive quantities, following their respective equations of state in eq.(\ref{eos}),  fluctuate about their mean values fixed by the constant reservoir potentials. Therefore, while by construction the reservoir potentials are constant the subsystem potentials can fluctuate. 
In this sense, the intensive variables are thermodynamic quantities, inasmuch as they are allowed to fluctuate, as opposed to the coupling strengths $J$ and $K$ which are fixed parameters.
Another instructive way of looking at intensive fluctuations would be think of small spatial inhomogeneities in $T,H$ or $D$ on the scale of the subsystem size. This would in turn induce an ``exchange" of extensive quantities between the system and the reservoir as per the Le Chatelier-Braun principle, in such a direction as to reduce the inhomogeneity, \cite{callen}. \footnote{ It might be mentioned here that for the magnetic systems the extensive quantities like the magnetization are unconstrainable, \cite{callen}. However, the formal fluctuation theory is equally applicable by consideration of a large control subsystem far removed from the system of interest and of size square root of the reservoir such that it renders the magnetization, etc globally conserved by absorbing non-conserving fluctuations, \cite{ruppmag}.} The inverse of the grand metric in eq.(\ref{fullmetric}) above gives the equilibrium fluctuations in the entropic intensive variables. We write it with raised indices, so that following Einstein summation $g^{ik} g_{kj}=\delta^1_j$
\begin{equation}
(g^{-1})_{ij}=g^{ij}=\langle\Delta x^i\,\Delta x^j\rangle
\label{fullmetricinv}
\end{equation}
From the above metric the fluctuations in the intensive variables $T,H,D$ can be found by obtaining their differential relation with the entropic intensive variables, \cite{sahay restricted},
\begin{eqnarray}
\Delta T &=& -\frac{1}{\beta^2}\,\Delta \beta\nonumber\\
\Delta H &=& \frac{\nu}{\beta^2}\,\Delta \beta - \frac{1}{\beta}\,\Delta \nu\nonumber\\
\Delta D &=& -\frac{\mu}{\beta^2}\,\Delta \beta + \frac{1}{\beta}\,\Delta \mu\nonumber\\
\label{deltaT}
\end{eqnarray}
Using eq.(\ref{fullmetricinv}) above the fluctuations in intensive variables can be calculated. The variance of $T$ is
\begin{equation}
\langle(\Delta\,T)^2\rangle = \frac{1}{\beta^4}\,g^{11}.
\label{delT}
\end{equation}
The variance of $H$ is
\begin{equation}
\langle(\Delta\,H)^2\rangle = \frac{\nu^2}{\beta^4}\,g^{11}-2\frac{\nu}{\beta^3}\,g^{12}+\frac{1}{\beta^2}\,g^{22}
\label{delH}
\end{equation}
and that of $D$ is
\begin{equation}
\langle(\Delta\,D)^2\rangle = \frac{\mu^2}{\beta^4}\,g^{11}-2\frac{\mu}{\beta^3}\,g^{13}+\frac{1}{\beta^2}\,g^{33}
\label{delD}
\end{equation}
Similarly we could obtain the cross moments like $\langle\Delta T\Delta H\rangle$, etc.

 The word ``grand" in the metric $g$ refers to the grand canonical ensemble wherein the system is in full thermodynamic contact with the reservoir, with all the extensive quantities fluctuating about the mean, \cite{sahay restricted}. On the other hand, we could also consider $restricted$ equilibrium fluctuations wherein the spin one system undergoes spontaneous motion only along limited directions in its state space. Let us explain this in more detail. Normally a thermodynamic constraint is understood in the context of some $process$ a system undergoes. For example, a gas in a piston could undergo an isobaric expansion so that a fixed pressure constrains the set of values of energy $E$, volume $V$ and particle number $N$ accessible to the equilibrium system. In its usual meaning, however, the isobaric constraint would imply a constant pressure only in the mean. This is understood by a consideration of the isobaric constraint $P(U,V,N)=constant$ as a hypersurface in the state space to which the equilibrium states of the gas would remain confined. However, at any instant the spontaneous equilibrium fluctuations could always take the system to nearby points away from the constant $P$-hypersurface. We could now impose a stronger ``$P$-constraint'' on the dynamics so that even the spontaneous thermodynamic fluctuations of the gas remain restricted to the $P$-hypersurface. We call it a $canonical$ constraint since it has the connotation of a canonical ensemble wherein while some variable are exchanged with the surrounding and hence can fluctuate about mean others are held strictly constant within ``walls''. Here, we expand the scope of the term to include restriction on fluctuations in the intensive quantities also, as discussed for the present example of a gas in a piston. 
 For the spin one model in principle one could restrict the full set fluctuations in as many ways as one could slice the state space. Some of them however, while mathematically admissible appear to be inaccessible or contrived experimentally. For example, on an $M$-surface wherein the magnetic moment fluctuations are suppressed, the fluctuations in $Q$ would have to peculiarly orchestrated. We shall return to this in a moment. 
 
As discussed in the preceding, the equilibrium fluctuations can be pictured equally well in the state space $\Gamma$ or the parameter pace $\mathcal{M}$. Since, as we shall see below, it is the intensive variables that are usually the independent variables in calculations for the spin one problem our reference space shall be the parameter space  $\mathcal{M}$ by default and unless otherwise stated the co-ordinates $x^i$ will refer to the entropic intensive variables $\beta,\nu,\mu$. The fluctuation metric for a thermodynamic system with a given canonical constraint is simply the metric induced on the corresponding hypersurface from the ambient grand metric. Let the co-ordinates for the parameter space be $x^i$ with $i=1..n$. Then a general canonical constraint is given by setting some function $f$ of the co-ordinates $x^i$ to be zero,
 \begin{equation}
f(x^1,x^2,..)=0
\end{equation} 
This defines a co-dimension one hypersurface in the state space which can be labelled by $n-1$ co-ordinates $y^a$ intrinsic to the hypersurface. We shall call it the $f$-surface. The components of the grand metric along the $f$-surface give the  projection metric in the $n$ dimensional parameter space,
\begin{equation}
g_{(f)\,ij}= g_{ij}-\,\frac{\partial_i\,f\,\partial_j\,f}{g^{kl}\partial_k f\,\partial_l f}
\label{projection}
\end{equation}
where $\partial_if$ is along the normal to the $f$-surface. In additon to the $n$-dimensional projection metric we can also write down the $n-1$-dimensional induced metric on the $f$-surface with its components along the $n-1$ intrinsic co-ordinates $y^a$ labelling the hypersurface,
 \begin{equation}
h_{(f)\,ab}=g_{(f)\,i\,j}\frac{\partial x^i}{\partial y^a}\,\frac{\partial x^j}{\partial y^b}=g_{\,i\,j}\frac{\partial x^i}{\partial y^a}\,\frac{\partial x^j}{\partial y^b}
\label{induced}
\end{equation}  
 The second equality stems from the fact that a tangent vector living on the $f$-surface is orthogonal to the gradient vector $\partial_if$. The two equivalent metrics in eq.(\ref{projection}) and eq.(\ref{induced}) provide complimentary information on fluctuations. Thus the induced metric $h_f$ gives the fluctuations in co-ordinates intrinsic to the $f$-surface,
\begin{equation}
h_{(f)\,ab}=\langle(\Delta Y_a\,\Delta Y_b)\rangle_f
\end{equation}
while
\begin{equation}
h_{(f)}^{\,ab}=\langle(\Delta y^a\,\Delta y^b)\rangle_f
\end{equation}
Here $Y_a $ labels suitable extensive quantities conjugate to the local coordinates defined on the hypersurface.
On the other hand, the projection metric directly reads off the $f$-constrained variance in the quantities defined on the full parameter space,
\begin{eqnarray}
g_{(f)\,\,ij}&=&\langle(\Delta X_i\,\Delta X_j)\rangle_f\nonumber\\
g_{f}^{\,ij}&=&\langle(\Delta x^i\,\Delta x^j)\rangle_f\nonumber\\
\label{gf}
\end{eqnarray}

We can immediately put to use the preceding results for the spin one model. Let us first write down the geometry obtained by restricting fluctuations in the intensive variable $D$. A constant $D$-hypersurface in the $\beta-\nu-\mu$ parameter space is the hyper-plane 
\begin{equation}
f(\beta,\nu,\mu)=\mu-\beta D = 0
\label{D plane}
\end{equation}
A suitable choice of local co-ordinates on the $D$-surface is 
\begin{equation}
y^1=\beta \,\,\,\,\mbox{and}\,\,\,\,   y^2=\nu. 
\label{D co-ords}
\end{equation}
so that global co-ordinates relate to them on the $D$-surface as
\begin{equation}
\beta(y^1,y^2)=y^1,\,\,\nu(y^1,y^2)= y^2,\,\,\mu(y^1,y^2) = D\,y^1
\label{intrinsic coord}
\end{equation}
 Using eq.(\ref{D plane})-eq.(\ref{intrinsic coord}) in eq.(\ref{induced}) the components of the induced metric $h_D$ on the $D$-surface are
\begin{eqnarray}
h_{(D)\,11}&=&g_{11}+ 2 D\, g_{13}+ D^2 g_{33}\nonumber\\
h_{(D)\,12}&=&g_{12}+ 2 D\, g_{32}\nonumber\\
h_{(D)\,22}&=&g_{22}
\label{induced D comps}
\end{eqnarray}
where the components $g_{ij}$ have their usual interpretation as variance of extensive quantities, see eq.(\ref{fullmetric}). Evidently, the results for the induced metric in eq.(\ref{induced D comps}) can be read off directly by simply setting $D$ to constant in eq.(\ref{dpsiU}) which gives
\begin{equation}
d\psi= -U_1\,d\beta - M\,d\nu\,\,\,\,\,\,\,(D \,\,\mbox{constant}),
\label{dPsiD}
\end{equation}
where $U_1=U+QD$ is an enthalpy like term. By taking second derivatives of $\psi$ with respect to $\beta$ and $\nu$ keeping $D$ constant, we obtain all the components of $h_D$. 

It can checked from eq.(\ref{gf}) and eq.(\ref{delD}) that on the $D$-surface the fluctuations in $Q$ are somewhat suppressed in comparison to the unconstrained case while the fluctuations in $M$ remain unaffected,
\begin{eqnarray}
\langle\,(\Delta Q)^2\,\rangle_D&=& \langle\,(\Delta Q)^2\,\rangle-\frac{1}{\beta^2\, \langle\,(\Delta D)^2\,\rangle }\nonumber\\
\langle\,(\Delta M)^2\,\rangle_D&=&\langle\,(\Delta M)^2\,\rangle
\label{fluc comp D}
\end{eqnarray}
Not surprisingly therefore, as we shall see in the next section, the scalar curvature on the $Q$-surface is aligned more with the fluctuations in the magnetization than the quadrupole moment.

Similarly, we can develop the geometry of the $H$-surface by projecting out the grand metric along the hypersurface defined by $\nu+\beta H=0$, where $H$ is a constant. Following the same procedure, the components of the induced metric $h_H$ along the local co-ordinates $y^1=\beta,y^2=\mu$ are,
\begin{eqnarray}
h_{(H)\,11}&=&g_{11}- 2 H g_{12}+ H^2 g_{22}\nonumber\\
h_{(H)\,12}&=&g_{13}- H g_{23}\nonumber\\
h_{(H)\,22}&=&g_{33}
\label{induced H comps}
\end{eqnarray}
Once again, the components of $h_H$ can be read off by setting $H$ to constant in eq.(\ref{dpsiU}),
\begin{equation}
d\psi= -U_2\,d\beta - Q\,d\mu\,\,\,\,\,\,\,(H \,\,\mbox{constant}),
\label{dPsiH}
\end{equation}
Here $U_2=U-MH$ is an enthalpy like term. One can check from eq.(\ref{gf}) and eq.(\ref{delH}) that on the $H$-hypersurface the $M$ fluctuations are weakened while the $Q$ fluctuations remain unrestricted.
\begin{eqnarray}
\langle\,(\Delta M)^2\,\rangle_H&=& \langle\,(\Delta M)^2\,\rangle-\frac{1}{\beta^2\, \langle\,(\Delta H)^2\,\rangle }\nonumber\\
\langle\,(\Delta Q)^2\,\rangle_H&=&\langle\,(\Delta Q)^2\,\rangle
\label{fluc comp H}
\end{eqnarray}
Evidently, as our subsequent discussion shall bear out, the scalar curvature on the $H$-surface encodes the fluctuations in $Q$ instead of $M$. 

Apart from the two preceding examples, we can also obtain hypersurface geometries corresponding to fixed $Q$ or fixed $M$. Using eq.(\ref{gf}) we can obtain, after some algebra, projection metrics on the surfaces $Q(\beta,\nu,\mu)=constant$ or $M(\beta,\nu,\mu)=constant$. Equivalently, we can take partial Legendre transforms of the entropy. Starting from eq.(\ref{dpsiU}) we obtain ,
 \begin{equation}
 d\psi_q=-U\,d\beta-M\,d\nu+\mu\,dQ
 \label{q const}
\end{equation}  
where $\psi_q=\psi+\mu\,Q$. Similarly, we obtain
\begin{equation}
 d\psi_m=-U\,d\beta-Q\,d\mu+\nu\,dM
 \label{m const}
\end{equation}  
where $\psi_m=\psi+\nu\,M$. The hypersurface metric can now be obtained by successive partial differentiation of the Massieu functions $\psi_q$ or $\psi_m$ by differentiating with respect to the entropic intensive variables while holding $Q$ or $M$ constant. However, as alluded to earlier, we shall not pursue these geometries here. The reason for this is that, owing to the kinematic coupling between $M$ and $Q$, freezing out fluctuations in one of them strongly constrains the fluctuations in the other. For example, using eq.(\ref{gf}), the fluctuations in $M$ at constant $Q$ are
\begin{equation}
\langle(\Delta\,M)^2\rangle_Q=\langle(\Delta\,M)^2\rangle-\frac{1}{||\partial Q||}\left(\frac{\partial Q}{\partial\nu}\right)^2
\label{mfluc qcons}
\end{equation}
where $||\partial Q||=g^{ij}\partial_iQ\partial_jQ $. On the other hand, in eq.(\ref{fluc comp D}) we see that the fluctuations in $M$ remain fully unconstrained on the $D$-surface.

We note that the scalar curvatures for all the two dimensional geometries can be obtained as sectional curvatures along their respective $f$-surfaces of the three dimensional Riemann curvature tensor of the grand metric. We shall call the scalar curvatures associated with the $H$-surface and the $D$-surface as the curvatures $R_q$ and $R_m$ respectively. The full three dimensional grand scalar curvature shall be denoted by $R_g$.

\section{Geometry of the one dimensional spin one model.  }
\label{onedgeo}

We now discuss the behaviour of the one dimensional spin one model through its state space geometry. Following our previous discussion, we shall be investigating the scalar curvatures associated with two geometries, namely that of the $D$-surface and the $H$-surface and the three dimensional scalar curvature of the full fluctuation metric.

The free energy per spin can be doubly differentiated via eq.(\ref{dPsiD}) to obtain the fluctuation metric on the $D$-surface and, similarly, via eq.(\ref{dPsiH}) to obtain the metric on the $H$-surface. We remind ourselves that there are two connections to be checked here. First, the $weak$ conjecture of Ruppeiner which relates the state space scalar curvature to the correlation length over a range not necessarily limited to the critical region, eq.(\ref{rcorr}). Second, the $strong$ conjecture equating the scalar curvature to the singular part of the free energy near criticality upto a universal constant of order unity, eq.(\ref{rpsi}). 

\begin{figure}[!t]
\begin{minipage}[b]{0.5\linewidth}
\centering
\includegraphics[width=2.in,height=1.8in]{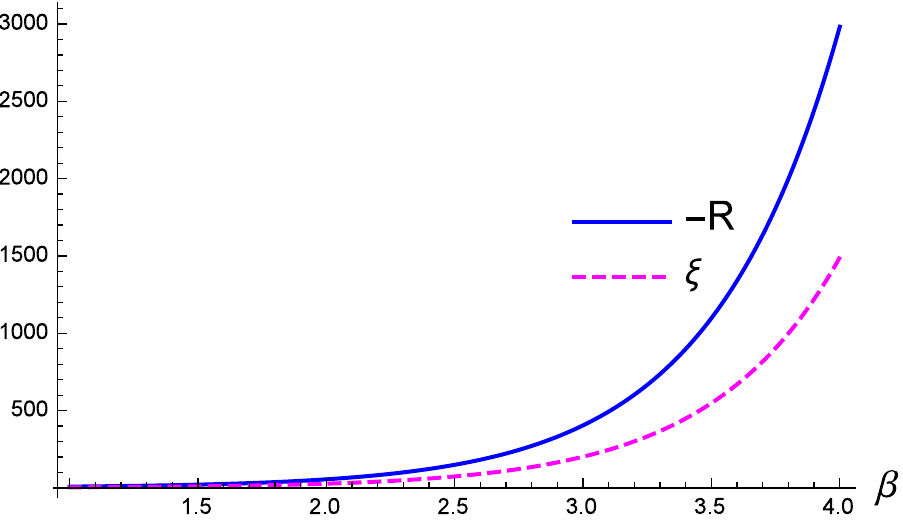}
\caption{\small{Temperature plots of $-R$ and $\xi$ for the Ising model in zero magnetic field with the coupling $J$ set to unity. Curvature goes as twice the correlation length at low temperatures.  }}
\label{ising1}
\end{minipage}
\hspace{0.6cm}
\begin{minipage}[b]{0.5\linewidth}
\centering
\includegraphics[width=2in,height=1.8in]{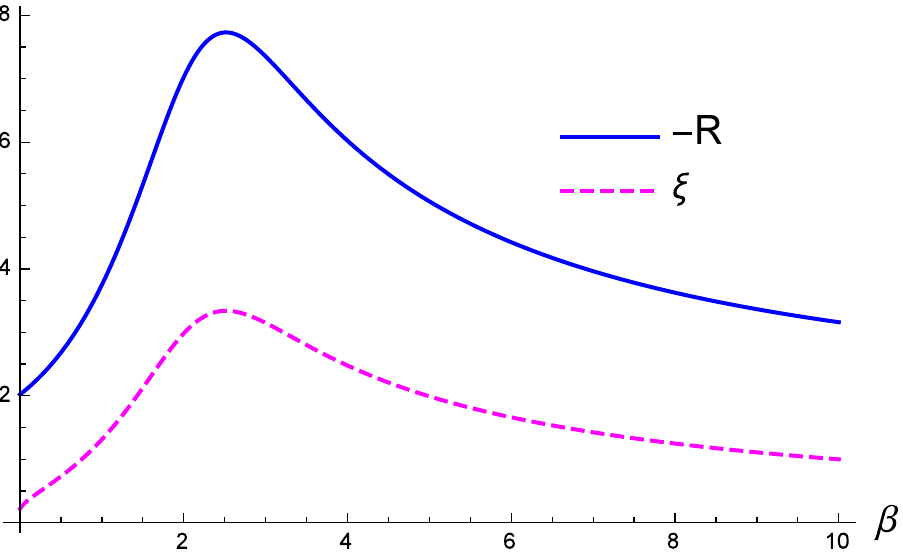}
\caption{\small{Temperature plots of $-R$ and $\xi$ for the Ising model in non-zero magnetic field $H=0.1$ with the coupling $J=1$. Curvature maintains a steady distance with the correlation length.  }}
\label{ising2}
\end{minipage}
\end{figure}

As a warm-up and to set the stage we do a lightning review of the geometry of the $1D$ ferromagnetic Ising model first discussed in \cite{ruppmag,mrugala}. The free energy and the correlation function can be obtained via the transfer matrix in the standard manner. The singular part of free energy near the pseudocritical point $T=0, H=0$ in the $\beta-H$ plane has the well known scaling form, \cite{pathria}
\begin{equation}
\psi_s= e^{-2\,J\,\beta}\left[1+\left(\frac{H\,\beta}{e^{-2\,J\,\beta}}\right)^2\right]^{1/2}
\label{freeising}
\end{equation}

and the correlation length as 
\begin{equation}
\xi^{-1}=\log \left(\frac{e^{2 \beta  H}+W+1}{e^{2 \beta  H}-W+1}\right)
\label{corrising}
\end{equation}
where $W^2=-2 e^{2 \beta  H}+e^{4 \beta  H}+4 e^{2 (H+2J)\beta}+1$.

while the scalar curvature is obtained as a simple expression, \cite{mrugala}
\begin{equation}
R=-\cosh (\beta\,H) \frac{1}{\sqrt{\exp (-4 J\,\beta )+\sinh ^2(\beta\,H)}}-1
\label{Rising}
\end{equation}

Therefore, it can be easily checked that $R$ follows the Ruppeiner equation at the pseudocritical point $H=0,\beta\to\infty$ with $\kappa=1$ and $-R\to2 \xi$ near the pseudocritial point. Satisfyingly, even for non zero magnetic field, which takes the system away from criticality, the curvature $R$ closely ``covers" the correlation length with the former asymptoting to $-2$ as $\xi$ decays to zero for large $\beta$. This is shown in fig.(\ref{ising1}) and fig.(\ref{ising2}). Thus, an exploration of the one dimensional Ising model reinforces both the strong and the weak ends of Ruppeiner's conjecture. Both the features will appear as a common theme in the geometry of one dimensional the spin one model to be discussed below. In addition, as we shall comment shortly, the Ruppeiner equation also helps straightforwardly obtain the scaling form of the free energy in eq.(\ref{freeising}), \cite{ruppspin}.

\subsection{$\bf{J>0}$}   
\label{jpositive}
    We first investigate the case of positive $J$. The zero temperature phase diagram for this case is given in fig.(\ref{krinA}). The lines marked $a$ in blue and $f$ in red are sites of pseudocriticality and two-phase coexistence while the point $b$ is a triple point as well as the pseudotrictritical point at zero temperature. On line $a$ the two order parameters $M$ and $Q$ have independent correlation lengths $\xi_1$ and $\xi_2$ as discussed earlier. While the fluctuations in the magnetization diverge in the limit $T=0$ on the $a$ line, the quadrupole fluctuations remain bounded and decay to zero on approaching $T=0$. Zero temperature value of $Q$ saturates to unity on either side of and also on the $a$ line.  From several scalar curvature vs. correlation length plots  we shall see that $\xi_1$ and $\xi_2$ correspond well with the scalar curvatures $R_m$ and $R_q$ respectively for much of the range of parameter values. There are variations to the theme however, possibly reflecting other attributes of $R$ apart from its association with the correlation length that require further investigation. On the other hand some of the unexpected behaviour could also be attributable to the geometry of the surfaces on which we have calculated the two curvatures.  We shall not make much attempt in this work to place some of the ``anomalous" behaviour of the curvatures within the context of underlying statistical interactions. Indeed, our analysis shall be necessarily limited at times, owing to the fact that a fundamental understanding of many features of $R$, especially its sign, is still a work in progress. In light of this it is important to undertake a somewhat detailed survey of the state space geometry in order to record patterns in variations, some of which could possibly have a bearing on a future analysis. In the following subsections we shall sort our observations of the scalar curvatures $R_q$, $R_m$ and $R_g$ for variations in parameters $H,D,J, \mbox{and } K$. Before doing so, we verify Ruppeiner's stronger conjecture near criticality and show that it directly suggests the form of the  scaling function associated with the singular free energy.
\subsubsection{Scaling form for the free energy}
\label{scaling}

We now analyze the free energy near (pseudo)criticality with the aim of connecting it to Ruppeiner's conjecture relating the singular part of the free energy to the scalar curvature, eq.(\ref{rpsi}). While the relation is not rigorously proved it has a strong physical justification and there is ample verification of it in several instances, see \cite{rupprev,ruppspin} and references therein.  One of the contributions of this paper is to add one more instance of its validity.
The conjecture is expected to be exact in the limit of the critical point where it will be called the Ruppeiner equation.  For the two dimensional parameter space the order unity universal constant $\kappa$ in eq.(\ref{rpsi}) works out to, \cite{ruppcrit}, 
\begin{equation}
\kappa=\frac{(b-1)(2b-a)}{a(a-1)}
\label{kappa}
\end{equation}
where $a$ and $b$ are the universal critical exponents in the scaling form of the singular free energy,
\begin{equation}
\psi_s(\beta,h) = n_1\, |t|^{a}\,Y(\frac{n_2 h }{|t|^{b}})
\label{scalingfree}
\end{equation}
In the expression for the singular free energy $t(\beta)$ is the reduced temperature, $h$ the ordering field and $Y(z)$ is the scaling function which depends only on the combination $z=h |t|^{-b}$. The constants $n_1$ and $n_2$ are non-universal and system dependent. For a system with a finite temperature critical point, like the Ising model in two or more dimensions the reduced temperature is $t=(\beta_c-\beta)/\beta_c$ and the ordering field is $h=H\beta$

For the three dimensional parameter space too the scalar curvature has been shown to be proportional to the singular part of the free energy at criticality, with the order unity proportionality constant $\kappa_3$ given as, \cite{ruppcrit}, 
\begin{equation}
\kappa_3=\frac{4a-4b-4c-4abc-a^2+4b^2+4c^2}{2(a-1)a}
\label{kappa3d}
\end{equation}
where the critical exponent $c$ is for the additional field $u$, apart from $t$ and the ordering field $h$, in the expression,
\begin{equation}
\psi_s(\beta,h,u) = n_1\, |t|^{a}\,Y(\frac{n_2 h }{|t|^{b}},\frac{n_3 u }{|t|^{c}})
\label{scalingfree3d}
\end{equation}
As it turns out, the two dimensional curvatures will be sufficient for our purpose of calculating the spin scaling functions in this section. We shall briefly revisit the constant $\kappa_3$ in eq.(\ref{kappa3d}) when we discuss the three dimensional curvature $R_g$ in later sections.

 Returning to the proportionality constant $\kappa$ in eq.(\ref{kappa}), it equals $1$ when either $a=b$ or $a=2\mbox{ and } b=0$. One dimensional models with short range interaction have critical point at zero temperature so that the the standard form of the reduced temperature will not apply.  Instead, for example in the case of the one dimensional ferromagnetic Ising model the reduced temperature can taken to be $t=e^{-p\,J\beta}$ where $p$ is any positive number. On setting $p=1$ the leading singular free energy becomes,\cite{fishnel}
\begin{equation}
\psi_s(h,t)=e^{-2J\beta}Y\left(\frac{h}{e^{2J\beta}}\right)\hspace{1cm}\mbox{(1-D Ising model)}
\label{scalingising}
\end{equation}
With $a=b=2$ above we get $\kappa=1$. Indeed, it would be the unity if $a$ and $b$ were both multiplied by any positive constant. In a recent work, \cite{ruppspin}, Ruppeiner worked out the form of the scaling function $Y(z)$ for the one dimensional Ising model by using the geometric Ruppeiner equation. This was done by first substituting the scaling form of the singular free energy $\psi_s$, given in eq.(\ref{scalingfree}), in the expression for $R$ which in two dimensional parameter space can be written as, \cite{rupprev}
\begin{equation}
R=-\frac{1}{2}\frac{\begin{vmatrix}
\psi_{,11}&\psi_{,12}&\psi_{,22}\\\psi_{,111}&\psi_{,112}&\psi_{,122}\\\psi_{,112}&\psi_{,122}&\psi_{,222}
\end{vmatrix}}{\begin{vmatrix}
\psi_{,11}&\psi_{,12}\\ \psi_{,21}&\psi_{,22}
\end{vmatrix}^2}
\label{ruppdet}
\end{equation}
and then equating the expression to $\kappa$ times inverse of $\psi$ in eq.(\ref{rpsi}). We can easily replace $\psi$ by $\psi_s$ in eq.(\ref{ruppdet}) because near criticality the leading behaviour of higher derivatives of the free energy is determined by its singular term. The resulting equation is a third order equation for the scaling function $Y(z)$ which contains only the variable $z$, the function $Y$ and its derivatives and the constants $a$ and $b$, with all dependencies on $\beta$ and $h$ dropping out. This is very much in keeping with the conjectured exactness of the Ruppeiner equation near criticality so that the only unknowns are the $universal$ exponents and the $universal$ scaling function. For the Ising model where the exponents $a=b$, the Ruppeiner equation further simplifies to
\begin{equation}
\frac{Y \left(Y \left(z Y^{(3)}+2 Y''\right)+z \left(z Y''^2-Y' \left(z Y^{(3)}+2 Y''\right)\right)\right)}{2 \left(Y-z Y'\right)^2 Y''}=\kappa
\label{scalingeqn}
\end{equation}
which becomes independent of the critical exponents.

A series expansion of $Y(z)$ in the equation above, keeping in mind that it must be an even function of $z$ for the ferromagnetic Ising model, produces a solution which matches exactly with its known form, namely
\begin{equation}
Y(z)=n_1\sqrt{1+(n_2z)^2},
\label{scalingfn}
\end{equation}
with $n_1=n_2=1$ for the Ising model. In addition, the series expansion of the left hand side of eq.(\ref{scalingeqn}) shows that the proportionality constant $\kappa$ is unity for the Ising model, which is consistent with eq.(\ref{kappa}) for $a=b$. While it is relatively straightforward to obtain the scaling function for the specific case of the one dimensional ferromagnetic Ising model ( see for example \cite{pathria}), in general it is a difficult problem in statistical mechanics as emphasized in \cite{ruppspin}. Indeed, it is remarkable that the scaling function can be obtained directly by a $thermodynamic$ equation. 

 As we shall discuss in the sequel, for the one dimensional spin one model too we have the exact relation $\kappa=1$ in the pseudocritical and the pseudotricritical limit. 
  Barring a few most of the checks are numerical. On approaching the $a$ line the curvature $R_m$ follows the Ruppeiner equation while $R_q$ remains finite, and on approaching the $f$ line and the pseudotricritical point $c$ both $R_m$ and $R_q$ follow the Ruppeiner equation. We recall here that $R_m$ and $R_q$ are both two dimensional sectional curvatures in a three dimensional parameter space, with $\mu,\nu$ and $\beta$ as one possible set of thermodynamic coordinates. $R_m$ is the intrinsic scalar curvature of the $D$-plane while $R_q$ is that of the $H$-plane. The scaling function of a free energy with three scaling fields would, in general, be $Y(z_1,z_2)$, with $z_1=h/t^a$ and $z_2=u/t^c$. Here, $h$ and $u$ are some linear combination of $\mu$, $\nu$ and $\beta$. For the case where both $h$ and $u$ are relevant so that $b,c>0$, the scaling function $Y$ becomes a genuine two parameter function and the Ruppeiner equation connecting the sectional curvatures with $\psi$ will result in a partial differential equation for $Y$. However, if one of the scaling fields, say $u$, is irrelevant with $c<0$ then the dependence on $u$ drops out near criticality and the scaling function becomes a one parameter function. Surely, with the $a$ and $f$ lines each being a locus of pseudocritical points the directions along the respective lines will not be relevant, so that the scaling functions near these lines will be one parameter functions $Y(z)$. Therefore, with a proper definition of the scaling field $h$, we expect the previously discussed differential equation for $Y$ to be valid in the vicinity of these lines. Clearly, on the $a$ line $Y(z)$ will be an even function of $z$ which should be proportional to $\nu=H\beta$ on the line. It is not a priori clear whether the scaling function is symmetric about the $f$ line. As our numerical studies seem to suggest, probably there are different scaling functions $Y_\pm(z)$ above and below the $f$ line. Nevertheless, we can treat each of the functions as symmetric in the scaling field $h$. At least our numerical checks seem to confirm this expectation in retrospect.

  Of the two possibilities consistent with $\kappa=1$, the values $a=2,b=0$ are easily ruled out because that would mean practically no dependence of the singular free energy term on the scaling field $h$ at very low temperatures, which is certainly not the case as can be easily checked. With $a=b$ equations (\ref{scalingeqn}) and (\ref{scalingfn}) apply to the spin one model as well. With more coupling terms in its Hamiltonian the reduced temperature $t$ for the spin one model is not always equal to $e^{-\beta J}$ but is of the more general form $e^{-\beta X}$ where $X$ is a linear combination of the coupling strengths in the Hamiltonian. Moreover, the form of $X$ changes depending on the relative strength of the coupling terms. The scaling form of the leading singular term in the neighbourhood of the $a$ line and in the vicinity of the $f$ lines outside the ``$f$-wings''(i.e, to the left  of $f$ line) can be written as
 \begin{eqnarray}
 \psi_s&=& n\,t\, Y\left(\frac{h}{n\,t}\right)\nonumber\\
 &=& n\,t\left[1+\left(\frac{h}{n \,t}\right)^2\right]^{\frac{1}{2}}\hspace{0.8cm}\mbox{( 1-D spin one model, outside $f$ wing)}
 \label{freespinone}
 \end{eqnarray}
 
 where the value of the constant $n$ and also the form of the reduced temperature $t$ depend upon the relative strength of the coupling terms. Note that our definition of the reduced temperature is such that the exponents $a=b=1$. In the vicinity of the pseudocritical $a$ line and the pseudotricritical point $c$ of fig.(\ref{krinA}) the ordering field $h=\beta H$, while around the upper and lower $f$ lines it is $h=(H-D+J+K)\beta$ and $h=(H+D-J-K)\beta$ respectively.
 
As mentioned earlier, the scaling form inside the ``$f$-wings" (to the right) is different. Our numerical checks suggest that it is of the following form,
  \begin{eqnarray}
 \psi_s
 &=& n\,t\left[1+\left(\frac{h}{n \,t}\right)^2\right]^{-\frac{1}{2}}\hspace{0.8cm}\mbox{( 1-D spin one model, inside $f$ wing)}\nonumber\\
 &\sim& \frac{n^2 t^2}{h}\hspace{1cm}(h>>t)
 \label{freespinonewing}
 \end{eqnarray}
 Indeed, this scaling form does not satisfy the differential equation (\ref{scalingeqn}) for $Y(z)$ with $\kappa=1$. Nonetheless, our checks seem to suggest this form of the scaling. We defer a detailed investigation of this observation to a future investigation.

 We first consider the $a$ line of fig.(\ref{krinA}) for which $D<J+K$. We recall that the $a$ line is a locus of pseudocritical points for the magnetization but not for the quadrupolar order. Satisfyingly, it is possible to separate the singular and regular parts of the free energy (or the Massieu function) near the $a$ line which we shall explain now. Recalling the expression for zero field Massieu function in eq.(\ref{psibegz}) we rewrite it keeping in mind that $D<J+K$ to get an expression of the form
 
 \begin{equation}
 \psi=(J+K-D)\beta+ \mbox{ln}\frac{1}{2}\left[1+e^{(D-J-K)\beta}+e^{-2J}+ \sqrt{1+w}\right]
 \label{freeex}
 \end{equation}
 where
\begin{equation}
w= e^{2(D-J-K)\beta}+8e^{(D-2J-2K)\beta}+e^{-4J}+2e^{-2J}-2e^{(D-3J-K)\beta}-2e^{(D-J-K)\beta}
\end{equation}
 The logarithmic term in  eq.(\ref{freeex}) goes to zero as $\beta\to\infty$. This is because with $D<J+K$ all the exponential terms within the logarithm, including those comprising $w$, are less than one and approach zero as $\beta$ becomes larger. Therefore, we can consider the first term in eq.(\ref{freeex}) as the regular part $\psi_r$ of the free energy near criticality and expect the singular part to be contained in the logarithm. The dominant part of the singular free energy must be that term in the logarithm which is the slowest to decay to zero as $\beta$ becomes larger. In order to filter out such terms we express the log term for large $\beta$ as follows,
\begin{equation}
\psi-\psi_r=\mbox{ln}\,\left[1+\frac{1}{2}e^{(D-J-K)\beta}+\frac{1}{2}e^{-2J}+\frac{1}{2}(\sqrt{1+w}-1)\right]
\label{psiminuspsireg}
\end{equation} 
and keep just the linear term in its expansion. However we must look at all the terms in the expansion of $\sqrt{1+w}$. This is because it leads to a systematic cancellation of all the terms containing  $e^{(D-J-K)\beta}$ or any of its higher powers which are inadvertently generated on expanding the square root term. The correlation length $\xi_1$ can also be analysed similarly on the $a$ line. In the limit of large $\beta$ it goes as the inverse of the singular free energy consistent with hyperscaling near the critical point. Finally, the leading singular free energy and the correlation length for magnetization fluctuations are obtained in the limit of the pseudocritical point as
\begin{eqnarray}
\psi_s &=&\,\frac{1}{2}\xi_1^{-1}\,= e^{-2J\beta} \hspace{1.9cm}(D<2K,D<J+K,H=0)\nonumber\\
&=& \frac{3}{4}\,\xi_1^{-1}=\,3\,e^{-2J\beta}\hspace{1.6cm}(D=2K,D<J+K,H=0)\nonumber\\
&=& \xi_1^{-1}\,\,\,\,\,=2\,e^{-(2J+2K-D)\beta} \hspace{1cm}(2K<D<J+K,H=0)
\label{freesingAline}
\end{eqnarray}
For all cases where $K\geq J$ the second and third possibilities will not arise. On the other hand, in the Blume-Capel limit $K=0$ we get different scaling behaviour for $D<0$, $D=0$ and $D>0$. The full scaling form for the free energy in the vicinity of the $a$ line is now easily written in the form of eq.(\ref{freespinone}) by choosing the values of $n=1,3$ and $2$ respectively for the above three cases in eq.(\ref{freesingAline}), the reduced temperature $t$ as the exponential term on the $r.h.s$ in each case and the ordering field as $h=H \beta$.

We now consider on the zero field line the point $D=J+K$ which is a triple as well as pseudotricritical point. As can be checked from its expression in eq.(\ref{freeex}) the free energy now goes to zero in the zero temperature limit. Its expression simplifies to
\begin{equation}
\psi=\ln \left[1+\frac{1}{2}e^{-2J\beta}+\frac{1}{2}e^{-2J\beta} \sqrt{1+8 e^{(3 J-K)\beta}}\right]
\label{freetric}
\end{equation}

 After a simple expansion of the logarithm we take the dominant part of the free energy as the one which is the slowest to decay as the pseudotricritical point is approached from non-zero temperatures. Depending on whether $K$ is less than, equal to or greater than $3J$ we have three cases,

\begin{eqnarray}
\psi_{s}&=&\sqrt{2}\,e^{-(J+K)\beta/2}\hspace{1cm}(K<3J,\,D=J+K,H=0)\nonumber\\
&=&2\,e^{-2J\beta}\hspace{2.2cm}(K=3J,\,D=J+K,H=0)\nonumber\\
&=&e^{-2J\beta}\hspace{2.4cm}(K>3J,\,D=J+K,H=0)
\label{freesingCpoint}
\end{eqnarray}
At the pseudotricritical point both the correlation functions $\xi_1$ and $\xi_2$ diverge with the following asymptotic relation to the singular free energy,
\begin{eqnarray}
\xi_1 &=&2\,\xi_2\,\,\,= \psi_s^{-1} \hspace{1.7cm}(K<3J,D=J+K,H=0)\nonumber\\
&=&  \xi_2\,\,\,\,\,\,\,=\frac{2}{3}\psi_s^{-1}\hspace{1.6cm}(K=3J,D=J+K,H=0)\nonumber\\
&=& 2\,\xi_2\,\,\,=\psi_s^{-1} \hspace{1.9cm}(K>3J,D=J+K,H=0)
\label{corrtric}
\end{eqnarray}
 Once again, the full scaling form for the free energy in the vicinity of the pseudotricritical point is expressed straightforwardly in the form of eq.(\ref{freespinone}) by choosing the values of $n=\sqrt{2},2$ and $1$ respectively for the three cases in eq.(\ref{freesingCpoint}), the reduced temperature $t$ as the exponential term on the $r.h.s$ in each case and the ordering field as $h=H \beta$.

Finally, we consider the $f$ line on which $D=|H|+J+K$. With non-zero $H$ it is easier to work with the numerical solutions for the eigenvalues of the transfer matrix. The $f$ line is a pseudocritical point for both the magnetic order and the quadrupolar order when approached from non-zero temperatures. For non zero field both the order parameters have the same correlation length $\xi_1$, \cite{krinsky}. Just as for the pseudotricritical point the free energy  goes to zero on the $f$ line and its singular part is double of the inverse correlation length
\begin{equation}
\psi_s=\frac{1}{2}\,\xi_1^{-1}=e^{-(J+K)\beta/2}\hspace{1cm}(D=|H|+J+K, \mbox{$f$ line})
\label{freesingFline}
\end{equation}

The full scaling form of the free energy can be written in a manner similar to the previous cases. We write the critical free energy explicitly for the region in the vicinity of the positive $f$ line, to its left, as
\begin{equation}
\psi_s =e^{-(J+K)\beta/2}\left[1+\left(\frac{H-D+J+K}{e^{-(J+K)\beta/2}}\right)^2\right]^{1/2}
\label{scalingformFline}
\end{equation}
and, to its right as
\begin{equation}
\psi_s =e^{-(J+K)\beta/2}\left[1+\left(\frac{H-D+J+K}{e^{-(J+K)\beta/2}}\right)^2\right]^{-1/2}
\label{scalingformFlinewing}
\end{equation}

\begin{figure}[t!]
\begin{minipage}[b]{0.3\linewidth}
\centering
\includegraphics[width=1.7in,height=1.2in]{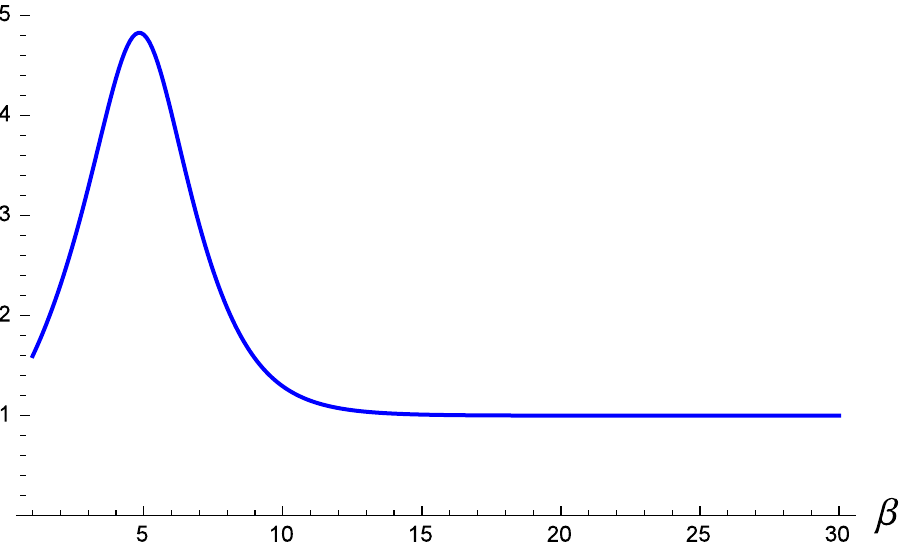}
\end{minipage}
\hspace{0.2cm}
\begin{minipage}[b]{0.3\linewidth}
\centering
\includegraphics[width=1.7in,height=1.2in]{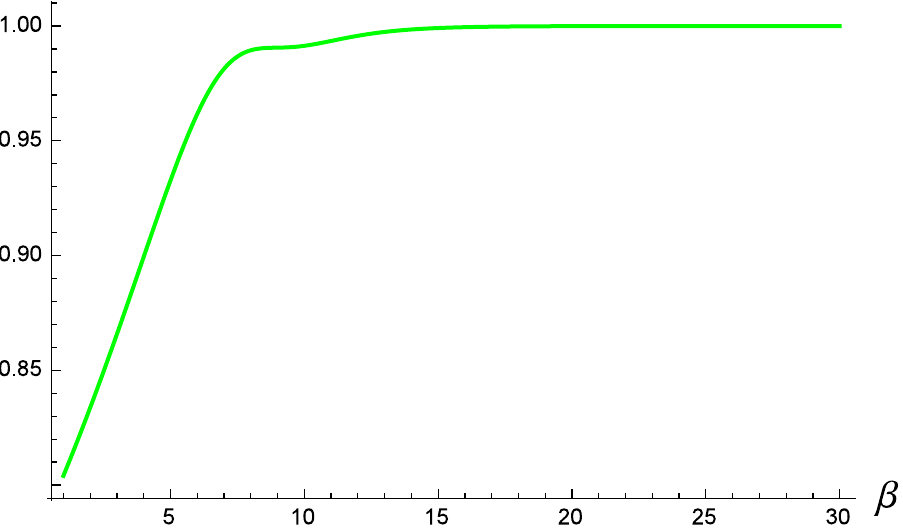}
\end{minipage}
\hspace{0.2cm}
\begin{minipage}[b]{0.3\linewidth}
\centering
\includegraphics[width=1.7in,height=1.2in]{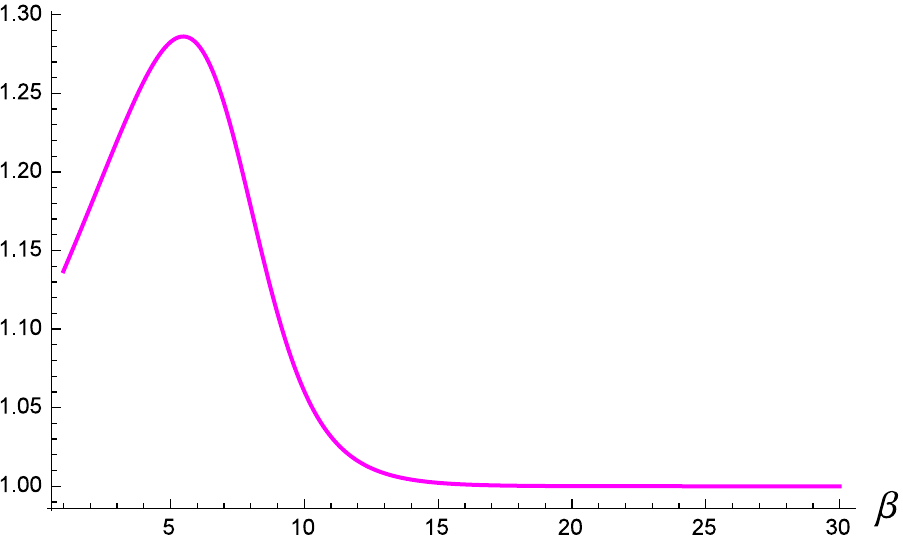}
\end{minipage}
\caption{\small{Plots of the ratio of the singular part of free energy to its three different scaling forms depending on different reference points. In all sub-figures the erference point $P$ at which the free energy is calculated has $J=0.3,K=0.2,H=0.01,D=0.49$ and it lies outside the $f$ ``wings". In $(a)$ the scaling form refers to the $a$ line , in $(b)$ the scaling form refers to the pseudotricritical point and in $(c)$ to the $f$ line. }}
\label{leop1}
\end{figure}
\begin{figure}[t!]

\begin{minipage}[b]{0.3\linewidth}
\centering
\includegraphics[width=1.7in,height=1.2in]{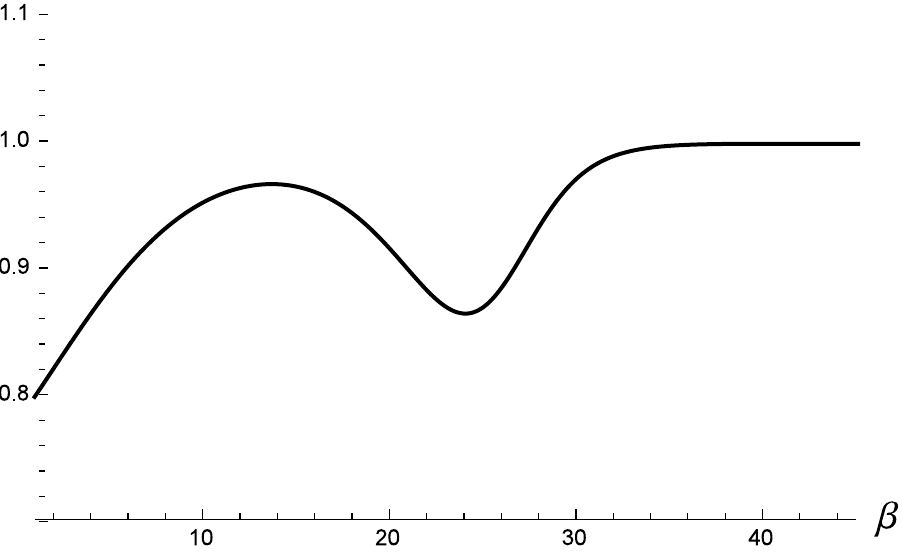}
\end{minipage}
\hspace{0.2cm}
\begin{minipage}[b]{0.3\linewidth}
\centering
\includegraphics[width=1.7in,height=1.2in]{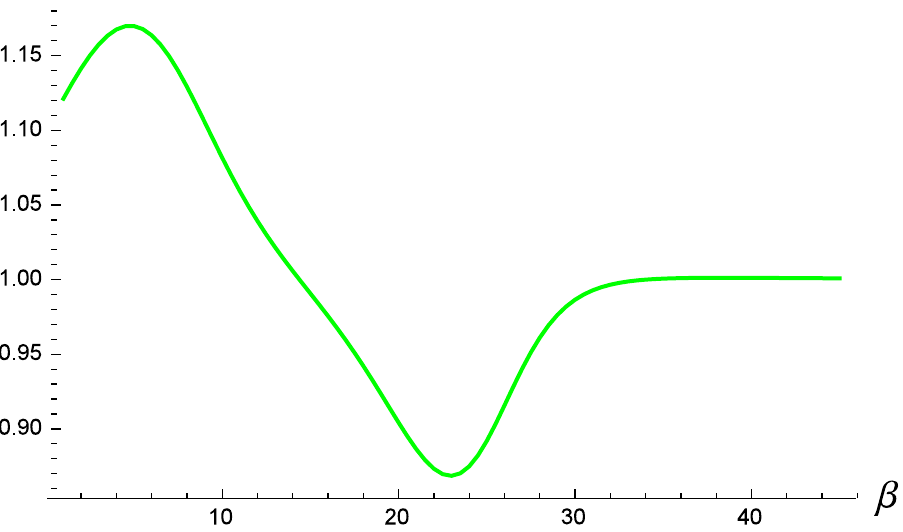}
\end{minipage}
\hspace{0.2cm}
\begin{minipage}[b]{0.3\linewidth}
\centering
\includegraphics[width=1.7in,height=1.2in]{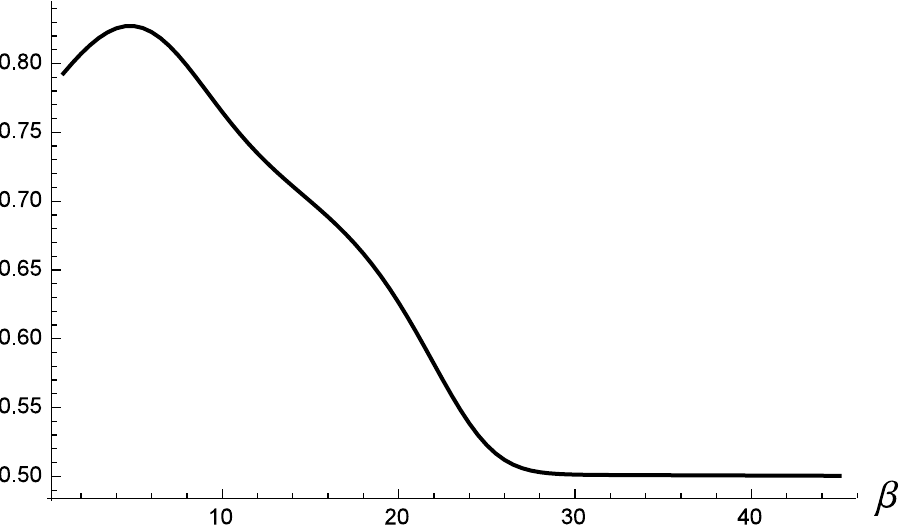}
\end{minipage}
\caption{\small{Plots of ratio of the free energy to its scaling forms for points inside the $f$ `wings'. The parameters $J=0.3$ and $K=0.2$ for all sub-figures. In $(a)$ the reference point at $D=0.5001$ and $H=0$ is referred to the vicinity of the pseudotrictritical point at $D=0.5,H=0$. In $(b)$ and $(c)$ the reference point at $D=0.5101$ and $H=0.01$ is referred, respectively, to the neighbourhood of the pseudocritical $f$-line and the pseudotricritical point.}}
\label{leop2}
\end{figure}
In fig.(\ref{leop1}) we present for a fixed point three plots of the ratio of the singular part of the free energy to its three different scaling forms obtained by a consideration of the point as existing in three intersecting neighbourhoods. The reference point which we shall label as $P$ lies slightly to the left of and above the pseudotricritical point $c$ (see fig.(\ref{krinA}) for reference). For all the three sub-figures all the parameter values are the same. In fig.\ref{leop1}(a) the point $P$ is considered in the neighbourhood of the $a$ line from a pseudocritical point directly below it. The singular free energy at $P$ is then taken to be $\psi-\psi_r$ as expressed in eq.(\ref{psiminuspsireg}) and the scaling form is ascertained from eqs.(\ref{freesingAline}) and (\ref{freespinone}). In fig.\ref{leop1}(b) the point $P$ is considered as belonging to the neighbourhood of the pseudotricritical point $c$ and the appropriate scaling form chosen accordingly following eqs.(\ref{freesingCpoint}) and (\ref{freespinone}). Finally, in fig.\ref{leop1}(c) the point $P$ is consireded in the vicinity of a pseudocritical point on the $f$ line at the same height as $P$, with the scaled free energy given in eq.(\ref{scalingformFline}). In all three cases the ratios approach unity at higher $\beta$ which is consistent with the respective scaling expressions for the free energy. 

  The uniformity in scaling discussed above is not present in the neighbourhood to the right of the pseudotricritical point and the $f$ line, in the region which is inside the $f$ ``wings" (refer to fig.(\ref{krinA})) . As stated earlier, within the wing region the scaling function follows eq.(\ref{freespinonewing}). In fig.\ref{leop2}(a) we plot for a point to the right of the $c$ point on the zero field line the ratio of the free energy to its scaling form, which is obtained from eq.(\ref{freesingCpoint}) and (\ref{freespinonewing}). Thus this point is considered as being in the vicinity of the pseudotricritical ponit $c$. In fig.\ref{leop2}(b) we consider a point in the wing region with a small non zero $H$. Considering it to be in the vicinity of a point on the $f$ line at the same height we plot the ratio of the free energy to its scaling form as given in eq.(\ref{scalingformFlinewing}). For both sub-figures (a) and (b) the ratio approaches unity thus affirming that these points do belong to their considered neighbourhoods. However, if now we consider the point of sub-figure \ref{leop2}.(b) to be in the neighbourhood of the pseudotricritical point and plot the ratio, as in fig.\ref{leop2}(c), we see that it does not converge to unity but to $0.5$. Therefore, our observations seem to suggest that only the zero field line constitutes a genuine neighborhood of the pseudotricritical point in the wing region. We can say that while the $a$ line, the $f$ line and the $c$ point share their neighbourhoods on the left hand side of the pseudotricritical point, the $f$ line and the $c$ point do not share their neighbourhoods on the right hand side. Presumably, this difference has to do with the presence of pseudocritical $a$ line on the left side but not on the right. In any case, the correctness of eq.(\ref{scalingformFlinewing}) stands verified.

Having utilized the strong conjecture of Ruppeiner to obtain the scaling function from thermodynamic geometry we turn our attention now to the investigation of the thermodynamic curvature vis-a-vis the correlation length, a relation which encompasses both the strong and the weak aspects of the conjecture. We shall examine the curvatures $R_q$ and $R_m$ for different parameter ranges and also comment on the three dimensional scalar curvature $R_g$. The main focus of our investigation will be the parameter values corresponding to the pseudocritical $a$ and $f$ lines and the pseudotricritical $c$ point and we shall observe the curvature, correlation length or the free energy as we approach these points from high temperatures. We shall also discuss the geometry when the magnetic field is non-zero.

\subsubsection{{\bf{Geometry of the zero field case, $H=0$}}}
\label{zeroH}
\paragraph{${\bf{J>K}}$}
\label{hzerojgreat}

 This case is the most relevant to the BEG model inasmuch as it refers to $\mbox{He}^3-\mbox{He}^4$ mixtures since for such mixtures $K$ is much less than $J$. Similarly, it is also relevant to the Blume Capel model for which $K=0$. The $a$ line, the pseudotricritical point $c$ and the part of of $x$ axis to its right in fig.(\ref{krinA}) constitute the zero field line. The $a$ line is a pseudocritical line with respect to $M$ fluctuations but not for $Q$ fluctuations.

 First we consider $R_q$ and $\xi_2$ on the whole of the zero field line. We note that we can obtain a closed form expression for $R_q$ by setting $H$ to zero in the thermodynamic metric defined on the constant $H$ surface, eq.(\ref{dPsiH}). While the zero field expression for $R_q$ is too large to be shown here, we obtain much smaller expressions for some special cases which we shall present on appropriate occasions. It is seen that for values of $D<K-J$ the magnitude of $R_q$ maintains an approximately constant distance with the correlation length $\xi_2$, with $R_q$ converging to $-1$ and the latter to zero as $T$ tends to zero. 
\begin{figure}[t!]
\begin{minipage}[b]{0.3\linewidth}
\centering
\includegraphics[width=1.7in,height=1.2in]{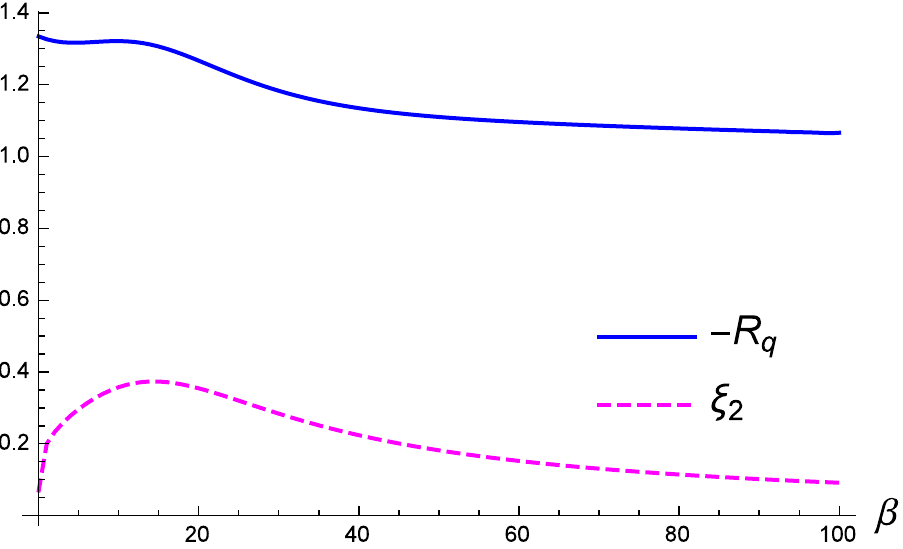}
\end{minipage}
\hspace{0.2cm}
\begin{minipage}[b]{0.3\linewidth}
\centering
\includegraphics[width=1.7in,height=1.2in]{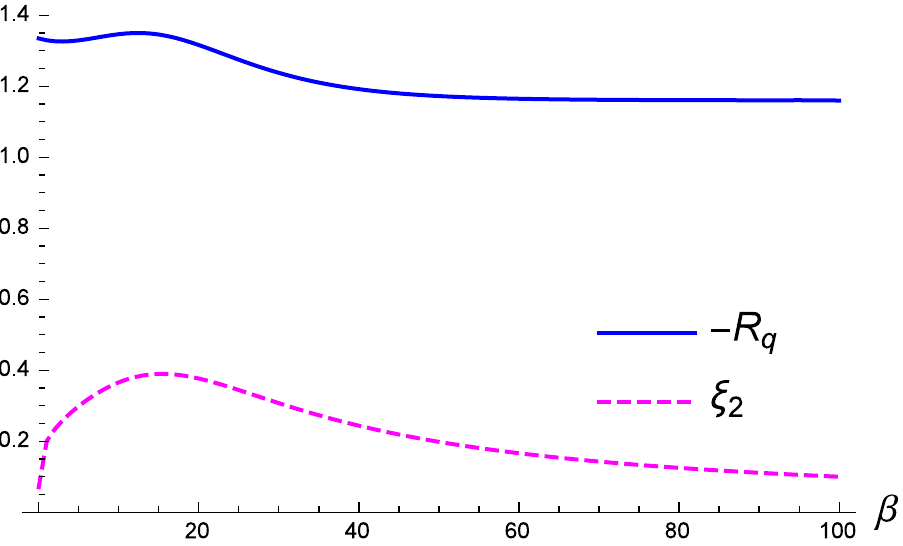}
\end{minipage}
\hspace{0.2cm}
\begin{minipage}[b]{0.3\linewidth}
\centering
\includegraphics[width=1.7in,height=1.2in]{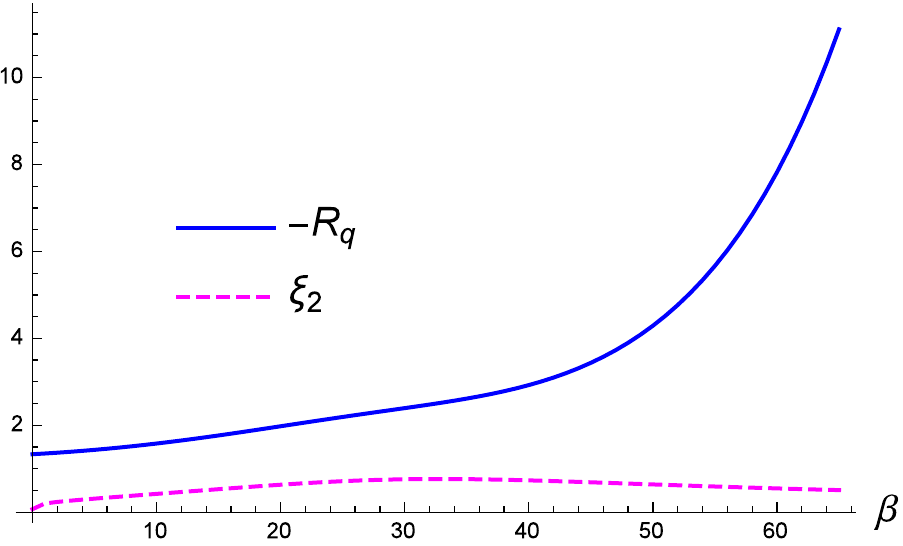}
\end{minipage}
\caption{\small{ Plots of $-R_q$ and $\xi_2$ against $\beta$ for different ranges of $D$ values, with $H=0$, $D<J+K$ and $K<J$. In all sub-figures $J=0.05,K=0.03$. In $(a)$ $D=-0.029<K-J$, in $(b)$ $D=-0.02=K-J$ and in $(c)$ $D=0.05>K-J$. In $(a)$ and $(b)$ the curvature remains almost parallel to the decaying correlation length while in $(c)$ it diverges away after running parallel for a while.}}
\label{norm1}
\end{figure}
 At $D=K-J$ the curvature $R_q$ converges to a value slightly less than $-1$ which depends only on the ratio $K/J$, with the minimum value of $-1.25$ in the Blume-Capel limit $K=0$ irrespective of the value of $J$. In figs. \ref{norm1}.(a)  and \ref{norm1}.(b) where we plot together the variation in the magnitude of $R_q$ and $\xi_2$ with respect to $\beta$, it can be seen that they maintain a separation of approximately one lattice unit for $D\le K-J$. We also note that the curvature-correlation length correspondence for the $H$-surface has no connection to criticality here. 
 
 For $K-J<D<K+J$ the curvature $R_q$ does not match with $\xi_2$ for the full range of temperature. While it parallels with $\xi_2$ for small values of $\beta$ (high temperature) it soon enough diverges away to more and more negative values. The negative divergence however is always less than or equal to the rate of divergence of $\xi_1$ or, equivalently, the negative divergence of $R_m$ as discussed below. While the apparently anomalous divergence could possibly provide some clue to the nature of underlying statistical interactions, we have already stated that we shall refrain from hypothesizing and defer such questions to a future investigation.   
 
\begin{figure}[t!]
\begin{minipage}[b]{0.3\linewidth}
\centering
\includegraphics[width=1.7in,height=1.2in]{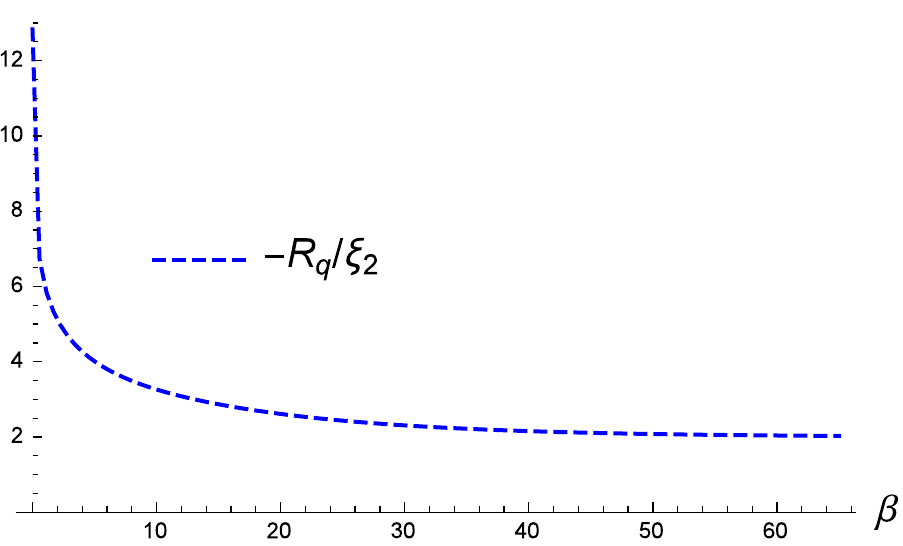}
\end{minipage}
\hspace{0.2cm}
\begin{minipage}[b]{0.3\linewidth}
\centering
\includegraphics[width=1.7in,height=1.2in]{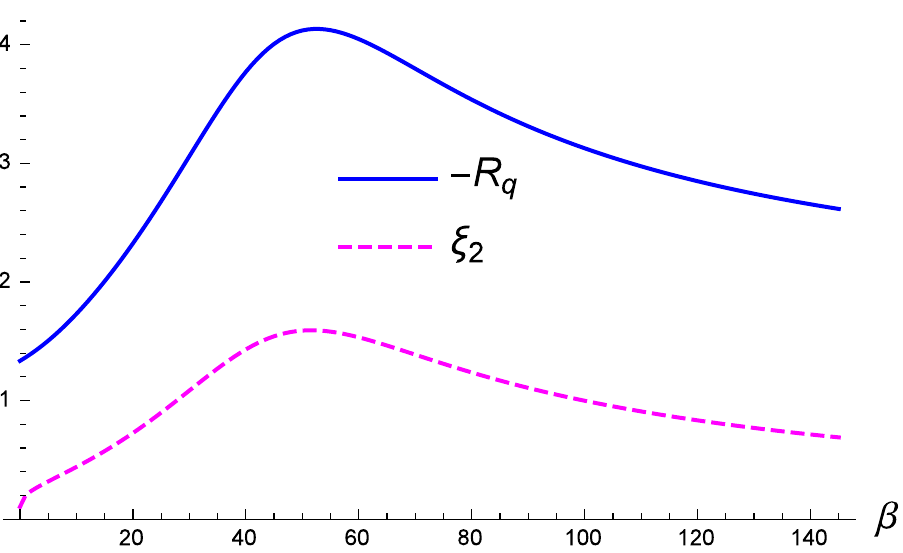}
\end{minipage}
\hspace{0.2cm}
\begin{minipage}[b]{0.3\linewidth}
\centering
\includegraphics[width=1.7in,height=1.2in]{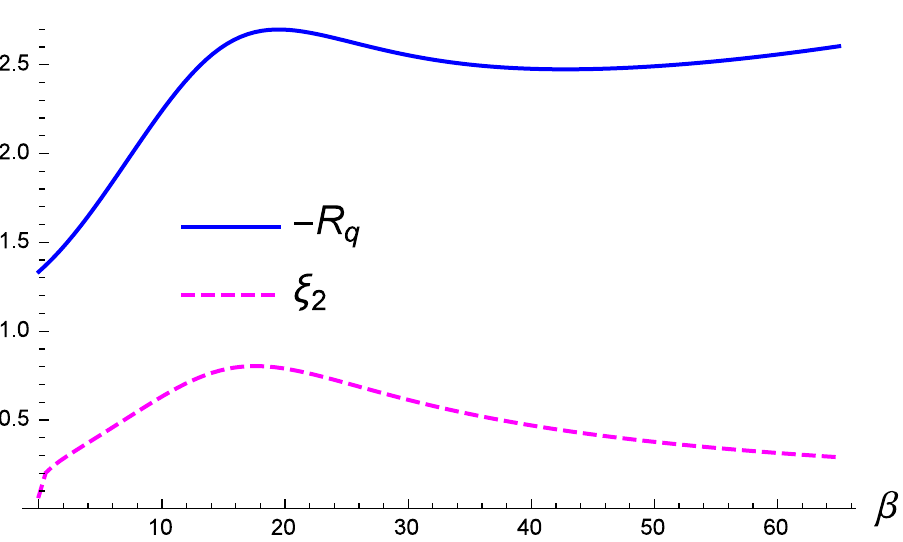}
\end{minipage}
\caption{\small{ Plots of $-R_q$ and $\xi_2$ against $\beta$ for different ranges of $D$ values, with $H=0$, $D\ge J+K$ and $K<J$. In all sub-figures $J=0.05,K=0.03$. In sub-figure $(a)$, with $D=0.08=K+J$(pseudotricritical point), the ratio $-R_q/\xi_2$ is plotted against the inverse temperature. In $(b)$ with $D=0.09<2J$ and in $(c)$ with $D=0.11>2J$ both $R_q$ and $\xi_2$ are plotted against $\beta$.}}
\label{norm2}
\end{figure}

  At the pseudotricritical point $D=J+K$, the negative of $R_q$ and $\xi_2$ both remain parallel and diverge asymptotically as $e^{(J+K)\beta/2}$, which is in line with eq.(\ref{freesingCpoint}). It can be checked that asymptotically $-R_q\sim 2\,\xi_2$, which is consistent with eq.(\ref{corrtric}) with $R_q\to \psi^{-1}$ in the pseudotricritical limit. Significantly, even at temperatures much farther from zero $R_q$ is already approximately twice $\xi_2$ as can be seen in fig. \ref{norm2}.(a).

 Beyond the tricritical point, for $J+K<D<2J$, the magnitude of $R_q$ and $\xi_2$ again run parallel to each other with the former converging to $-2$ and the latter decaying to zero as $T$ tends to zero. This is shown in  fig.\ref{norm2}(b). For $D=2J$ the curvature $R_q$ converges to a value between $-2$ and $-4$ with the latter value fixed for the Blume-Capel limit irrespective of $J$. For $D>2J$ $R_q$ behaves anomalously and diverges to negative infinity after following the correlation length $\xi_2$  for small $\beta$, as shown in fig.\ref{norm2}(c). The asymptotic rate of divergence is $e^{(D-2J)\beta}$.

In the Blume-Capel limit $K=0$ the tricritical point expression for $R_q$ simplifies to
\begin{equation}
R_q=\frac{\mathcal{N}_1}{\mathcal{D}_1 }\hspace{0.2in} (D=J,H=0,K=0)
\label{Rqbc}
\end{equation}
where $\mathcal{N}_1$ and $\mathcal{D}_1$ are given in eq.(\ref{A1}) of the Appendix. 
The high temperature limit of $R_q$ can be expressed in terms of the ratio $\alpha=K/J$, irrespective of $D$,
\begin{equation}
R_q=\frac{-4 \alpha ^4+27 \alpha ^2-243}{2 \left(\alpha ^2+9\right)^2} \hspace{1in}(\beta\to0)
\end{equation}
For the Blume-Capel case $R_q\to-3/2$ as $T\to\infty$.
\begin{figure}[t!]
\begin{minipage}[b]{0.3\linewidth}
\centering
\includegraphics[width=1.7in,height=1.2in]{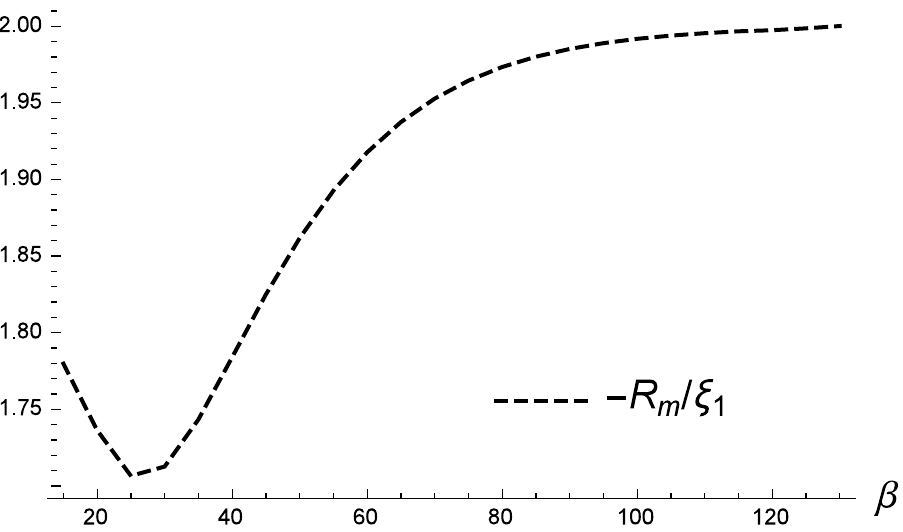}
\end{minipage}
\hspace{0.2cm}
\begin{minipage}[b]{0.3\linewidth}
\centering
\includegraphics[width=1.7in,height=1.2in]{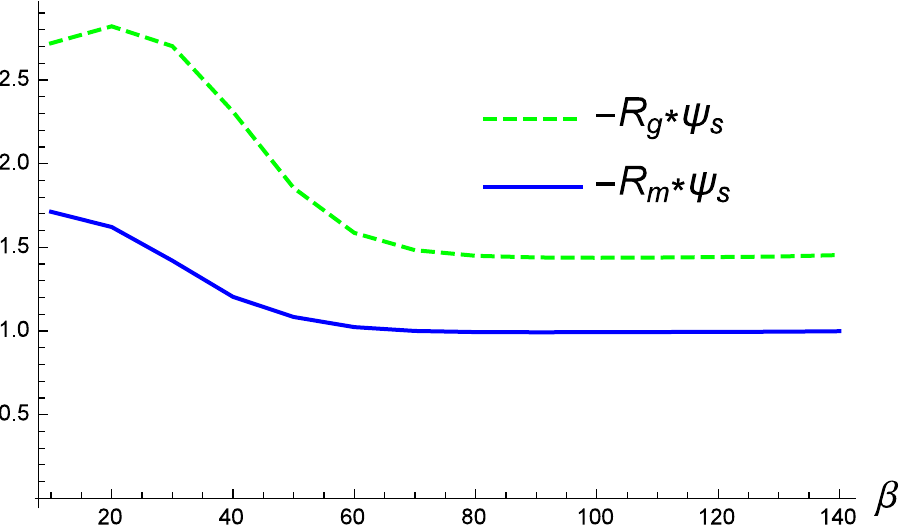}
\end{minipage}
\hspace{0.2cm}
\begin{minipage}[b]{0.3\linewidth}
\centering
\includegraphics[width=1.7in,height=1.2in]{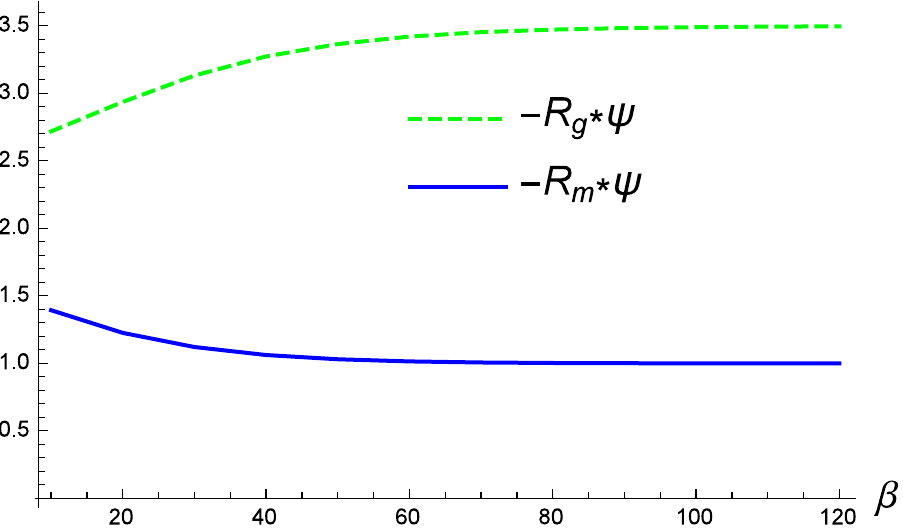}
\end{minipage}
\caption{\small{In all sub-figures $J=0.05,K=0.04$.$(a)$ Plot of $R_m/\xi_1$ vs. $\beta$ with $D=0.02$.$(b)$ Plots of $-R_g\,(\psi-\psi_r)$ and $-R_m\,(\psi-\psi_r)$ vs. $\beta$ for $D=0.082$. $(c)$ Similar as $(b)$ but with  $D=J+K=0.09$.}}
\label{norm3}
\end{figure}

We now consider $R_m$ and the three dimensional scalar curvature $R_g$ on the zero field line. Closed form expressions are not available for these curvatures since the eigenvalues of the transfer matrix with non-zero $H$ are to be obtained by solving a cubic equation.  The curvature $R_m$  the correlation length $\xi_1$ run parallel to each other upto the pseudotricritical point, i.e, for $D\leq J+K$. The ratio of the curvature to correlation length is fully consistent with eq.(\ref{freesingAline}) for all range of values of the parameter $D$, with $R_m\to \psi_s^{-1}$ towards zero temperature. In fig.(\ref{norm3})(a) we observe that the ratio $R_m/\xi_1$ soon enough approaches the value of $2$ which, for the chosen parameter values in the figure, is consistent with eq.(\ref{freesingAline}). The full curvature $R_g$ is seen to closely parallel the curvature $R_m$. In fig. (\ref{norm3})(b) and (c) we plot together the products of $R_m$ and  $R_g$ with $\psi_s$ for different values of $D$, with the latter sub-figure at the pseudotricritical value $D=J+K$. We note here that while the two dimensional scalar curvature $R_m$ is seen to follow the Ruppeiner equation with $\kappa=1$ consistent with eq.(\ref{kappa})same it is not the same with the three dimensional curvature $R_g$. While $R_g$ still follows the Ruppeiner equation at criticality in that it goes as the inverse of singular free energy, the proportionality constant $\kappa$ does not appear to be a universal number, though it is still of order one. There are patterns to the variation. For example, at the pseudotricritical point for $K<3J$ the constant $\kappa_3=3.5$ while for $K\ge 3J$ it is equal to $3$. In this work we do not pursue issues of whether or not the proportionality $\kappa_3$ follows eq.(\ref{kappa3d}) and shall take it up in the future.
Beyond the pseudotricritical point while the correlation length $\xi_1$ decays to zero at low temperatures $R_m$ does not match with $\xi_1$ and diverges in the negative direction.

\paragraph{${\bf{J=K}}$}
\label{hzerojeq}
\begin{figure}[t!]
\begin{minipage}[b]{0.3\linewidth}
\centering
\includegraphics[width=1.7in,height=1.2in]{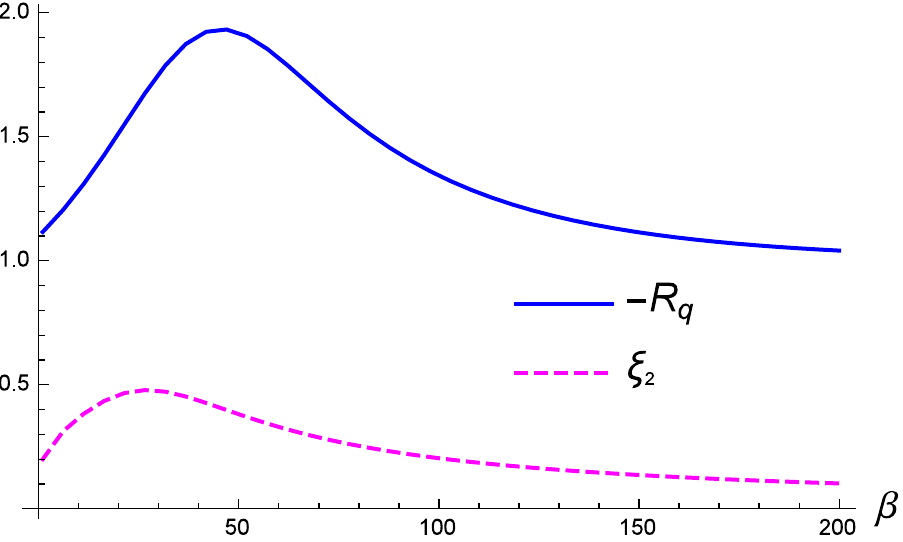}
\end{minipage}
\hspace{0.2cm}
\begin{minipage}[b]{0.3\linewidth}
\centering
\includegraphics[width=1.7in,height=1.2in]{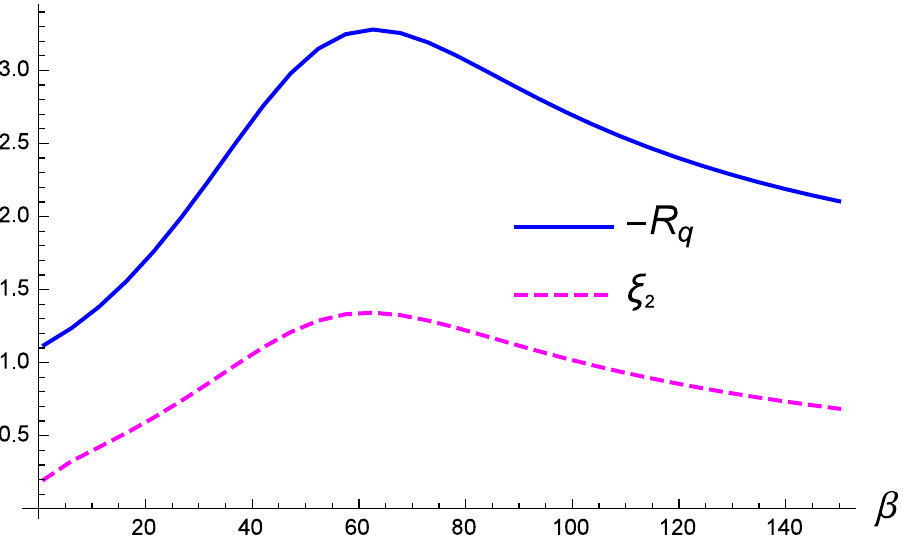}
\end{minipage}
\hspace{0.2cm}
\begin{minipage}[b]{0.3\linewidth}
\centering
\includegraphics[width=1.7in,height=1.2in]{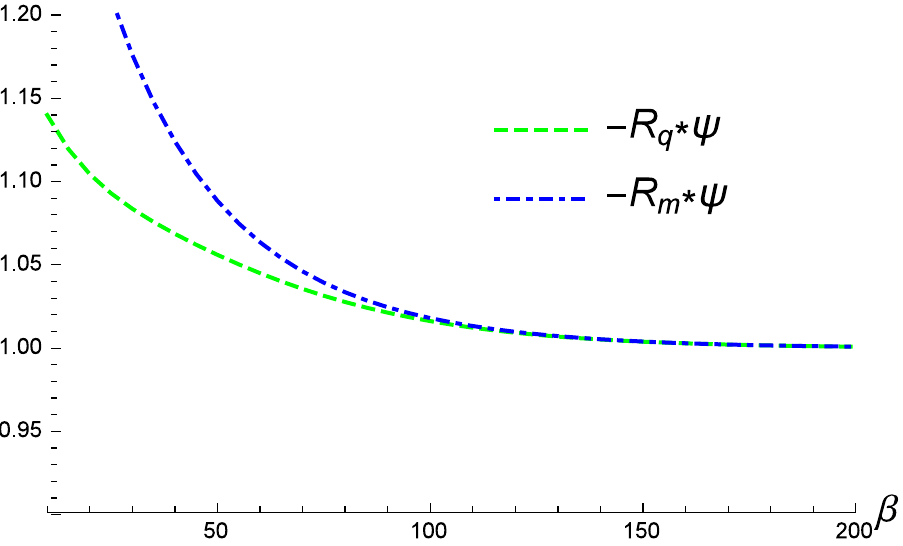}
\end{minipage}
\caption{\small{In all sub-figures $J=K=0.03$ and $H=0$. $(a)$ and $(b)$ are plots of $R_q$ and $\xi_2$ vs. $\beta$ for  $D=0.05$ and $D=0.07$ respectively . In $(c)$ plots of $-R_q\,\psi$ and $-R_m\,\psi$ are shown at the pseudotrictritical point for $D=J+K=0.06$}}
\label{norm4}
\end{figure}
For the case when the spin coupling strength is equal to the quadrupole coupling, the curvature $R_q$ shows a very good correspondence with $\xi_2$ for all values of $D$. For $D<J+K$ the curvature $R_q$ runs parallel to $\xi_2$ and asymptotes to $-1$ as the correlation length tends to zero for low temperatures. Similarly, for $D>J+K$ it asymptotes to $-3/2$ as $\xi_2$ tends to zero. For the pseudotricritical point $D=2J$, $-R_q$ and $\xi_2$ diverge to infinity as $e^{J\,\beta}$ with $R_q \to 2\,\xi_2$ at low temperatures, consistent with eq.(\ref{freesingCpoint}) and eq.(\ref{corrtric}). 
For $J=K$ the zero field expression for $R_q$ reduces considerably and it can be expressed as the fraction
\begin{equation}
R=\frac{\mathcal{N}_2}{\mathcal{D}_2}
\end{equation}
where the numerator and denominator are given in the appendix, eq.(\ref{A2}).

$R_m$ corresponds well with $\xi_1$ upto the pseudotricritical point. For $D<2J$ the curvature $R_m$ equals twice $\xi_1$ for low temperatures while for the pseudotricritical point $D=2J$ it equals $\xi_1$ at low temperatures in line with eq.(\ref{corrtric}). For $D<2J$ the asymptotic divergence of $\xi_1$ and $R_m$ is $e^{2J\beta}$, while at the pseudotricritical point it is $e^{J\beta}$, all of which follows eq.(\ref{freesingCpoint}). Once again, beyond the pseudotricritical point $R_m$ does not match with $\xi_1$ and diverges in the negative direction in a manner similar to the $J>K$ case.

In fig.\ref{norm4}(a) the curvature $R_q$ and $\xi_2$ are plotted for a value of $D<J+K$ and in fig.\ref{norm4}.(b) for $D>J+K$. For both cases there is a very good correspondence between curvature and correlation length as discussed above. In fig.\ref{norm4}.(c) the product of the magnitude of $R_m$ and $R_q$ with the free energy is plotted for the pseudotricritical value of $D$. The two curvatures are seen to approach each other and the product quickly reaches a value of $\kappa=1$ following the Ruppeiner equation. Furthermore, for the three dimensional curvature the product $R_g \psi$ approaches $3.5$ as mentioned earlier.

\paragraph{${\bf{J<K}}$}
\label{hzerojless}

\begin{figure}[t!]
\begin{minipage}[b]{0.3\linewidth}
\centering
\includegraphics[width=1.7in,height=1.2in]{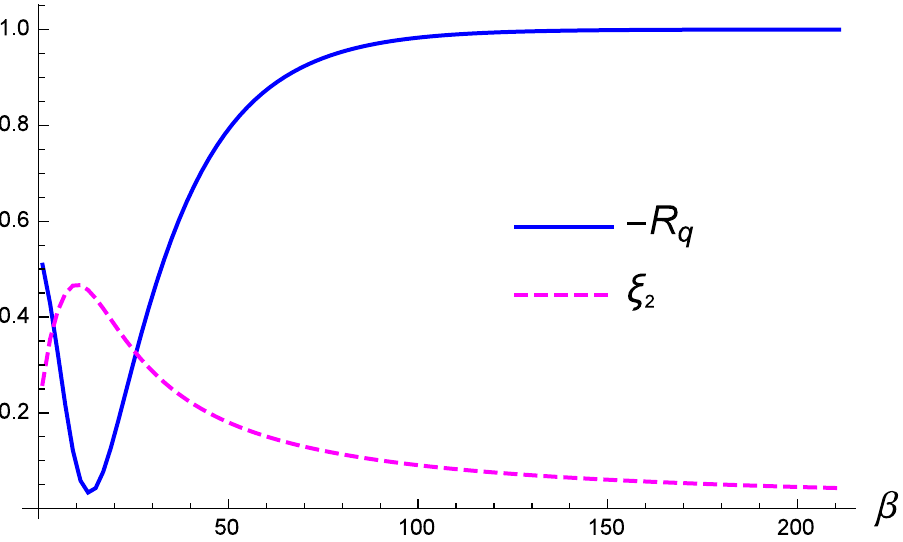}
\end{minipage}
\hspace{0.2cm}
\begin{minipage}[b]{0.3\linewidth}
\centering
\includegraphics[width=1.7in,height=1.2in]{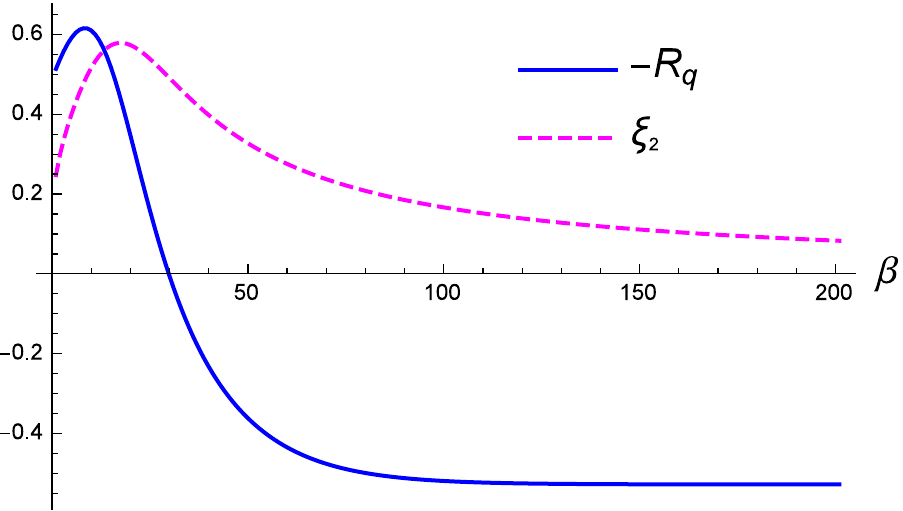}
\end{minipage}
\hspace{0.2cm}
\begin{minipage}[b]{0.3\linewidth}
\centering
\includegraphics[width=1.7in,height=1.2in]{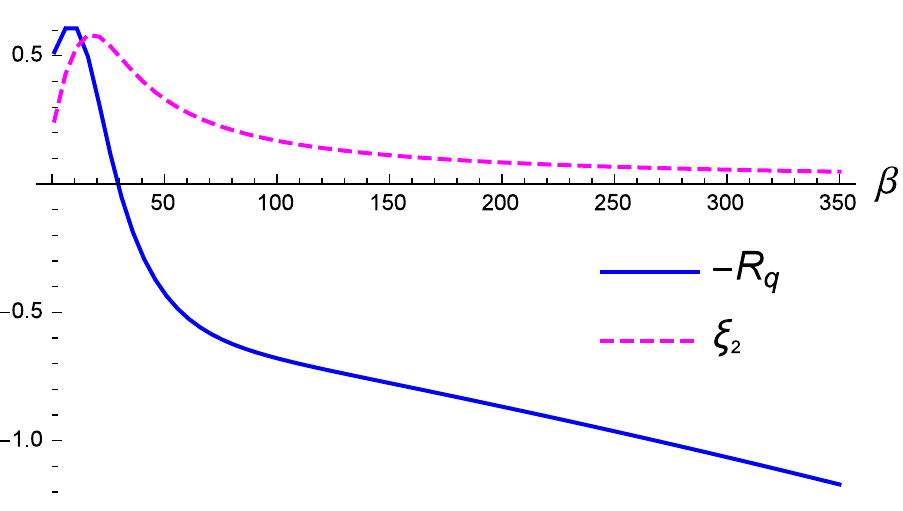}
\end{minipage}
\caption{\small{ Plot of $R_q$ and $\xi_2$ $vs.$ $\beta$ with $(a)$ $J=0.03,K=0.1,D=0.02<K-J,H=0$, $(b$) with  $J=0.03,K=0.08,D=K-J=0.05,H=0$ and $(c)$  with $J=0.03,K=0.08,D=0.051>K-J,H=0$. }}
\label{norm5}
\end{figure}

We now survey the geometry for the case when the quadrupole coupling strength is greater than the dipole coupling. For values of $D\leq K-J$ the curvature $R_q$ asymptotes to $-1$ towards zero temperature. However, contrary to the previous cases, $-R_q$ does not rise to more positive values around the regions of maxima of $\xi_2$. Instead, the negative curvature dips towards the negative direction around the maxima of $\xi_2$, only to asymptote to $1$ at lower temperatures. The nearer $D$ gets to $K-J$  the longer is the range of temperature for which the negative of curvature dips and stays to less positive values, even crossing the $y$-axis for $D$ close to $K-J$,  before finally rising to positive values and approaching $1$ for very low temperatures. In fig.\ref{norm5}.(a) is shown a plot of $-R_q$ and $\xi_2$ $vs.$ $\beta$ for for $D<K-J$. There is a pronounced dip in the negative curvature before if finally converges to one.  At $D=K-J$ the negative curvature $-R_q$ no longer rises to positive values at low temperatures but asymptotes to a negative value depending only on the ratio $K/J$. For example for $K/J=8/3$ the curvature $-R_q$ goes to $-19/36$ at zero temperature, as shown in fig. \ref{norm5}.(b). For $K-J<D<K+J$, the $-R_q$ does not follow the correlation length $\xi_2$ and diverges to negative infinity as $e^{(D-K+J)\beta}$, as shown in fig. \ref{norm5}.(c). This is always slower than the divergence to infinity of $\xi_1$ which goes as $e^{2J\beta}$ following eq.(\ref{corr}).

At the pseudotricritical point $D=J+K$ the curvature $R_q$ behaves differently compared to the previous two cases $J>K$ and $J=K$.  Here, it is also important to note the values of the quadrupole moment $Q$ and its fluctuation $\langle(\Delta Q)^2\rangle$ the reason for which will soon be clear. For $K\le3J$ $R_q$ diverges to negative infinity as $e^{(J+K)\beta/2}$ and follows the Ruppeiner equation with $\kappa=1$.
For $K<3J$ the quadrupole moment $Q$ saturates to $1/2$ at zero temperature and its fluctuation $\langle(\Delta Q)^2\rangle$ diverges to infinity as $e^{(J+K)\beta/2}$. At $K=3J$ the quadrupole moment equals to $2/3$ in the zero temperature limit. For $3J<K<5K$ the curvature continues to diverge to minus infinity at the same rate as the correlation length $\xi_2$ which now diverges as $e^{2J\beta}$. However, now at low temperatures -$R_q\,\psi=a$  where $a$ is less than $1$ and it reduces substantially as $K$ approaches $5J$.
\begin{figure}[t!]
\begin{minipage}[b]{0.3\linewidth}
\centering
\includegraphics[width=1.7in,height=1.2in]{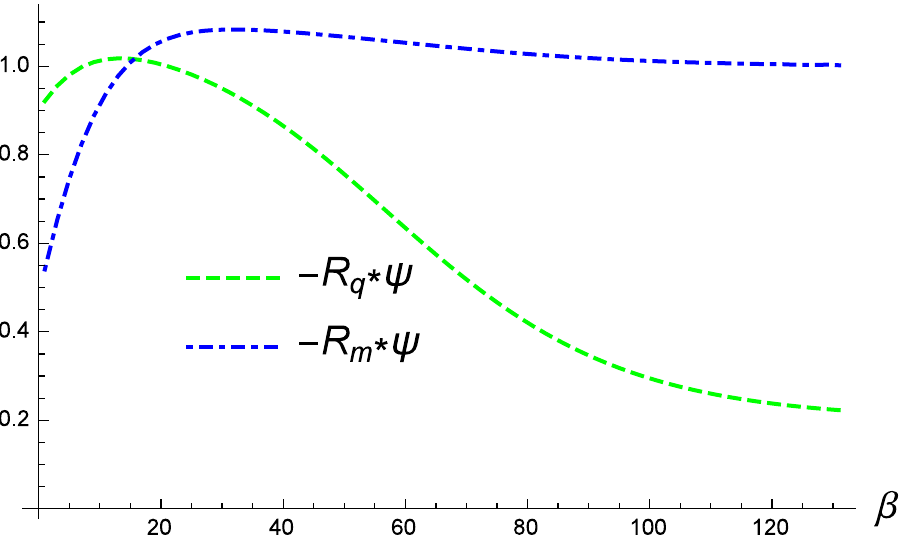}
\end{minipage}
\hspace{0.2cm}
\begin{minipage}[b]{0.3\linewidth}
\centering
\includegraphics[width=1.7in,height=1.2in]{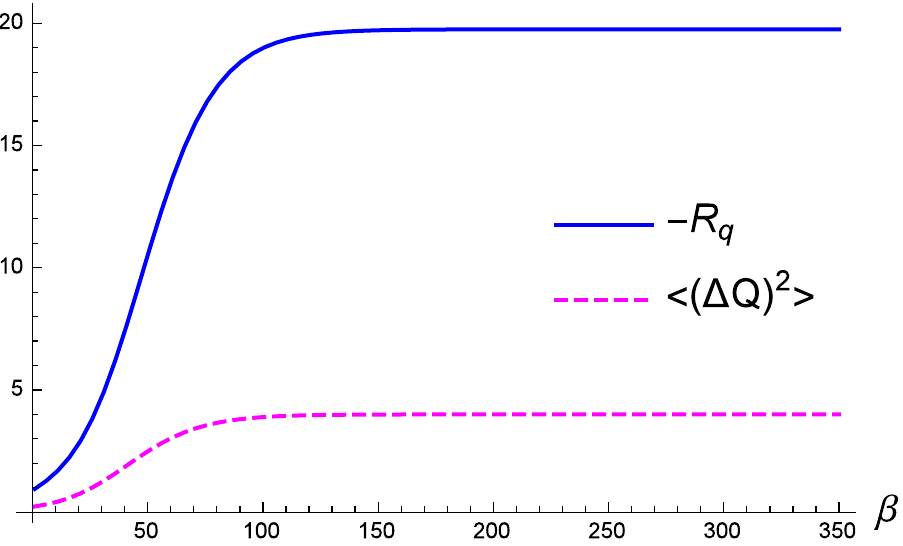}
\end{minipage}
\hspace{0.2cm}
\begin{minipage}[b]{0.3\linewidth}
\centering
\includegraphics[width=1.7in,height=1.2in]{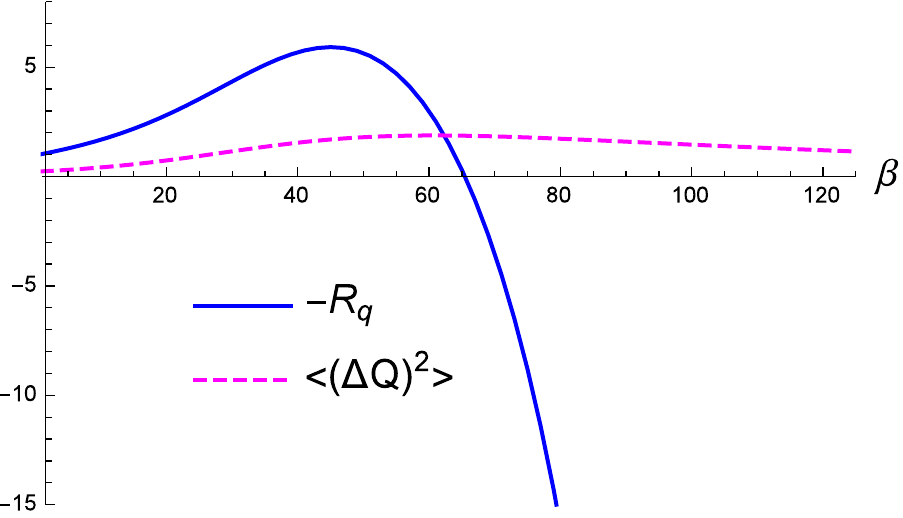}
\end{minipage}
\caption{\small{Zero field plots with $D$ at pseudocritical values $D=J+K$ and $K>J$. $(a)$ Plot of the products of the singular free energy $\psi$ with $R_q$ and $R_m$, for $J=0.03,K=0.14<5J,D=0.17$. $(b)$ Plot of $R_q$ and $<(\Delta Q)^2>$ $vs$. $\beta$ with $(b)$ $J=0.03,K=0.15=5J,D=0.18$ and $(c)$ $J=0.03,K=0.16>5J,D=0.19$  }}
\label{norm6}
\end{figure}
 For example as $K$ increases from $0.149$ to $0.1499$, for $J=0.03$, $a$ decreases almost ten times from $\frac{299}{14400}$ to $\frac{2999}{1440000}$. Thus, while $R_q$ still diverges to negative infinity in this parameter range, it does not strictly follow the Ruppeiner equation. This can be seen in fig. \ref{norm6}(a) where $R_m$ clearly follows the Ruppeiner equation but not $R_q$.  Meanwhile, for all values of $K>3J$ the quadrupole moment tends to $1$ and its fluctuation goes as $e^{(5J-K)\beta}$. Therefore, as $K$ approaches $5J$ the divergence in quadrupole fluctuations kepdf flattening. This might appear curious since the quadrupole-quadrupole correlaton length $\xi_2$ grows steadily as $e^{2J\beta}$. At $K=5J$, the quadrupole fluctuation completely flattens and approaches the fixed value of $4$. Exactly paralleling this situation the curvature $R_q$ too stops diverging and asymptotes to a fixed negative value of $-79/4$ at $K=5J$, as seen in fig. \ref{norm6}(b). The correlation length continues to diverge at the same rate as before. For $K>5J$ the quadrupole fluctuation decays to zero in the zero temperature limit. Now the curvature $R_q$ diverges to $positive$ infinity at the same rate as the correlation length, namely as $e^{2J\beta}$. This is shown in fig. \ref{norm6}(c). This is interesting because in general the scalar curvature diverges to negative infinity at the critical point. 
 For $D>J+K$ the curvature $R_q$ diverges to positive infinity as $e^{(D-2J)\beta}$ after briefly following the correlation length.

The apparent inconsistency of the quadrupole correlation length diverging to infinity and at the same time the quadrupole fluctuations decaying to zero at the pseudotricritical point is understood by examining the expression for the quadrupole fluctuation,\cite{krinsky},
\begin{equation}
\frac{\partial^2\psi}{\partial \mu^2}=\frac{1}{N}\langle(\ \sum_{i=1}^N \Delta Q_i)^2\rangle=\frac{Q(1-Q)}{1-\lambda_3/\lambda_1}
\end{equation}
so that while at the pseudotricritical point $\lambda_3\to\lambda_1$ which implies $\xi_2\to\infty$, the numerator could compete with the zero in the denominator if $Q\to1$. This is more pronounced for $K>5J$ when the quadrupole fluctuation now decays to zero while the correlation length continues to diverge as before. It is significant that the curvature $R_q$ encodes this peculiar behaviour of quadrupole fluctuations.

The curvature $R_m$ follows the Ruppeiner equation for all $K>J$ and for all $D\le K+J$. The asymptotic ratio between $R_m$ and the singular free energy (and also with $\xi_1$) is reached much before criticality thus again emphasizing the weak conjecture of Ruppeiner. Beyond the pseudotricritical point $R_m$ behaves exactly as in the previous cases in that it diverges at zero temperature to negative infinity even as $\xi_1$ decays to zero.

\subsubsection{Geometry of non-zero H: the $f$ line and beyond }
\label{hnonzero}

For non-zero magnetic field there is only one correlation length, namely $\xi_1$, for both the spin and the quadrupole fluctuations. We note that we cannot obtain closed form expressions for the scalar curvatures in this case. The non zero field region comprises the $f$ line and its vicinity, the regions to its left and right and the vicinity of the pseudotricritical point. We shall survey each region one by one.len1a

\begin{figure}[!h]
\begin{minipage}[b]{0.5\linewidth}
\centering
\includegraphics[width=2.3in,height=1.8in]{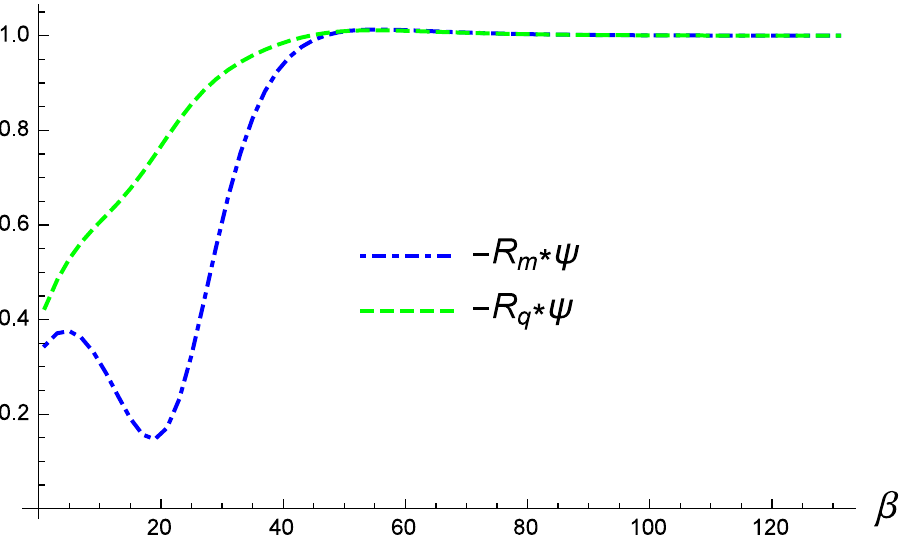}
\caption{\small{Plot of the products $R_m\psi$ and $R_q\psi$ $vs.$ $\beta$ on the $f$-line. The parameters $J=0.03,K=0.1,H=0.01,D=H+J+K=0.14$.}}
\label{len1}
\end{minipage}
\hspace{0.6cm}
\begin{minipage}[b]{0.5\linewidth}
\centering
\includegraphics[width=2.3in,height=1.8in]{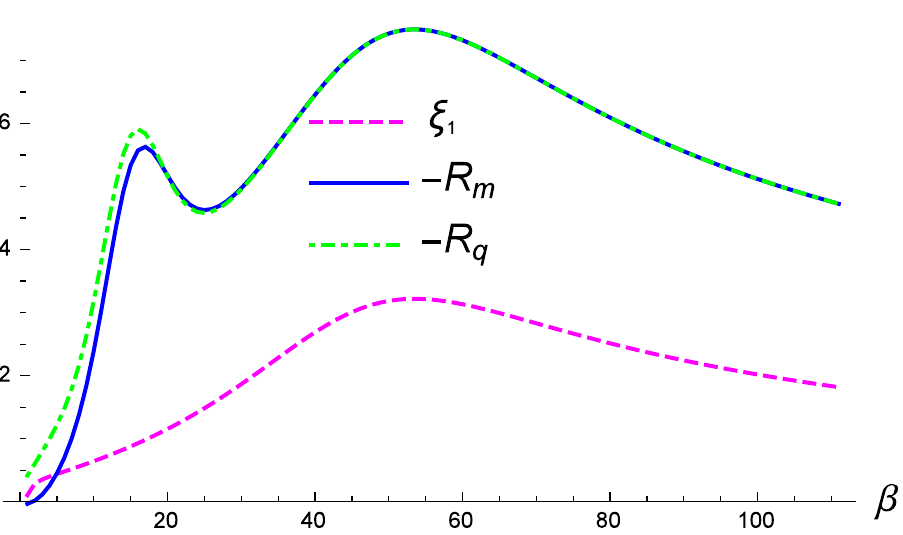}
\caption{\small{A representative plot of $R_m$, $R_q$ and $\xi_1$ in the $f$ band, with parameter values $J=0.1,K=0.09,D=0.45$ and $H=0.25$. The hump in the curvatures and their substantial overlap is noticeable.}}
\label{len2}
\end{minipage}
\end{figure}

As stated earlier, the $f$ line is the boundary at $T=0$ such that to its right, the energetically preferred state is the one with $S=0$ for all spins while to its left the lowest energy state has all the spins aligned with $S=1$ for positive magnetic field. On the $f$ line both the curvatures $R_q$ and $R_m$ diverge and follow the Ruppeiner equation with $\kappa=1$ as shown in fig.(\ref{len1}). As is apparent from the figure, both curvatures merge into each other at low temperatures which is consistent with the fact there is only one correlation length for non-zero field. The scaling of scalar curvatures is $e^{(J+K)\beta/2}$ which follows from eq.(\ref{freesingFline}). 
\begin{figure}[t!]
\centering
\includegraphics[width=3.5in,height=2.5in]{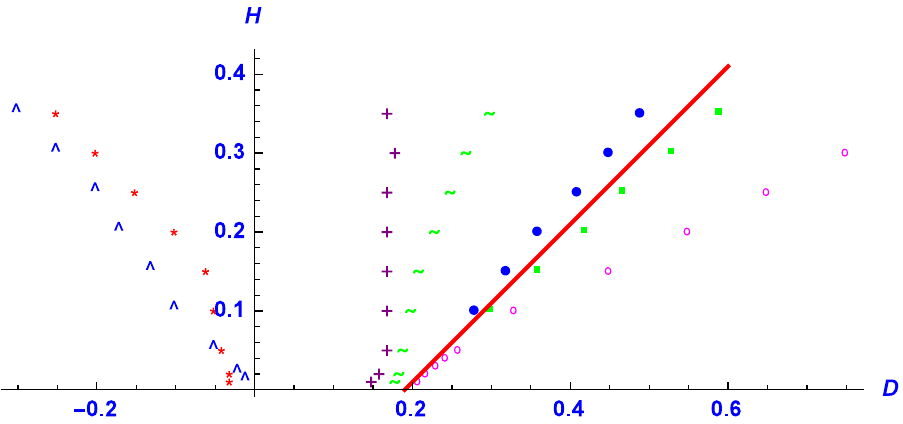}
\caption {A diagram showing trends in the curvatures $R_m$ and $R_q$ in the $H-D$ plane with $J=0.1,K=0.09$. See text for figure description.}
\label{curvtrends}
\end{figure}

In order to explore further the geometry of non-zero $H$ we refer to fig.(\ref{curvtrends}) which records trends in the curvatures $R_m$ and $R_q$. The values of parameters $J$ and $K$ are fixed as $0.1$ and $0.09$ respectively. Turning our attention first to the vicinity of the red coloured $f$ line we notice that starting from small non-zero values of $H$ onwards there is a wedge shaped band around the $f$ line, bordered by blue dots on the left and green squares on the right. Within this band it is observed that the two curvatures $R_m$ and $R_q$ overlap very strongly at low temperatures and moreover they show a very good correspondence with the decaying correlation length. Further, the band is characterized by  an additional high temperature local maximum (or a ``hump") in both the $R_m\, vs.\, \beta$ and $R_q\,vs.\,\beta$ plots, see fig.(\ref{len2}). While in this work we shall not seek to understand the physical reason for the hump within the ``f-band" we comment that the geometry retains a possible memory of the pseudocritical $f$ line within the band. We note that smaller values of $H$ (about $0.05$ in the figure), namely for points close to the pseudotricritical point, the f-band shrinks to nearly zero  and the ``hump" feature disappears too, suggesting a unique neighbourhood of the tricritical point.

Moving leftwards from the band towards decreasing values of $D$, one encounters a succession of points at which one or the other of $R_q$ or $R_m$ diverge to positive or negative values. Of course, in all this the correlation length $\xi_1$ always decays to zero. While the overlap is excellent within the $f$-band the two curvatures begin to separate on moving leftwards but still follow each other till the curve marked by green coloured ``tilde" shaped symbols is reached. From this point leftwards $R_q$ diverges to negative infinity until on the extreme left the curve marked by the blue coloured ``caret" symbols is reached from whence the curvature $R_q$ begins to diverge towards positive infinity. On the other hand, on crossing the curve marked by the purple coloured ``plus" symbols the curvature $R_m$ diverges towards positive infinity followed by its divergence towards negative infinity on crossing the curve marked by red coloured ``asterisk" symbols. There occur two regions in which the direction of divergence of both the curvatures is in opposite direction. The first region being between the ``plus"  and the ``asterisk" in which $R_q$ and $R_m$ diverge towards negative and positive infinity respectively. The other region being beyond the curve marked by caret symbols to the left of which the direction of divergence of both the curvatures are reversed. 
To the right of $f$-band, for $D$ greater than but close enough to $J+K+H$,  $R_m, R_q$ and the correlation length $\xi_1$ have a good match till the curve marked by pink coloured ``open circles" is reached. To the right of the pink curve both the curvatures diverge to negative infinity. The three dimensional curvature $R_g$ more or less follows the curvature $R_m$ so we do not describe it separately here. 

Admittedly, we do not understand the physics behind the several sign changes of the curvatures here but given that the sign of the scalar curvature is commonly associated with the underlying statistical interactions it is important to record patterns in its variation. Further analysis shall be the subject of a future investigation.

\subsection{ J=0 case, the one dimensional Griffiths model}
\label{jzero}

The zero temperature phase diagram for the one-dimensional Griffiths model, namely the BEG model in the limit $J=0$, was described earlier in sec.(\ref{onedgeo}) and represented in fig.(\ref{krinB}). It was mentioned that the whole of the $x$-axis is non-critical in this case and that the $f$ line with $D=K+H$ is the locus of pseudocritical points at zero temperature. The absence of the spin coupling strength $J$ simplifies the transfer matrix in eq.(\ref{transfer}) enough to render closed form expressions for two eigenvalues, with the third eigenvalue identically zero.

\begin{eqnarray}
\lambda_{\pm} &=& \frac{1}{2} e^{ -(D+ H)\beta} \left(e^{( D+ H)\beta}+e^{(2H+ K)\beta}+e^{\beta  K} \pm \sqrt{W}\right)
\label{eigenjz}
\end{eqnarray}
where
\begin{equation}
W= (e^{( D+ H)\beta}+e^{(2 H+ K)\beta}+e^{K\beta})^2+4 (e^{(D+ H)\beta}+e^{(D+3 H)\beta}-e^{(D+H+K)\beta}-e^{(D+3 H+K)\beta})
\label{Wjz}
\end{equation}
From the above equation we can calculate the free energy $\psi$ as log $( \lambda_+)$ and the correlation length $\xi$ as $1/\,\mbox{log}\,(\lambda_+/\lambda_-)$. Starting with the free energy closed form expressions can be obtained for $R_m$, $R_q$ and $R_g$. We note that while the numerical values of the curvatures in this section can be obtained from those of the previous section on setting $J=0$, closed form expressions were not available earlier for non-zero $H$. We can obtain the scaling form of the free energy on the $f$ line by obtaining its approximate expression for large values of $\beta$ 
\begin{equation}
\psi \underset{\beta\to\infty}{\to}  \frac{1}{2}e^{-2\,H\,\beta}+\frac{1}{2}\sqrt{4 e^{2 \beta  H}+4 e^{4 \beta  H}+e^{\beta  K}}
\label{freejzero}
\end{equation}
The leading singular term of the free energy and its relation to the correlation length on the $f$ line with $D=H+K$ can now be obtained as
\begin{eqnarray}
\psi_s &=&\,\xi^{-1}\,= e^{-2\,H\,\beta} \hspace{4cm}(H<K/4)\nonumber\\
&=& \frac{1+\sqrt{5}}{2\sqrt{5}}\,\xi^{-1}=\frac{1+\sqrt{5}}{2}\,e^{-2H\beta}\hspace{1.6cm}(H=K/4)\nonumber\\
&=& \frac{1}{2}\xi^{-1}\,\,\,\,\,=2\,e^{-K\beta/2} \hspace{3.2cm}(H>K/4)
\label{freesingfzline}
\end{eqnarray}
  It would also be instructive to know about the thermal fluctuations in magnetization and quadrupole moment in view of their bearing upon the geometry in a manner analogous to sec.(\ref{hzerojless}). We present the quadrupole fluctuations on the $f$ line $D=K+H$ for different ranges of $H$ values,
 \begin{eqnarray}
\langle (\Delta Q)^2\rangle &=& 2\,e^{(6 H-K)\beta} \hspace{3.5cm}(H<K/4)\nonumber\\
&=&\frac{2}{5\sqrt{5}}\,e^{2 H\beta}\hspace{3.5cm}(H=K/4)\nonumber\\
&=& \frac{1}{4}\,e^{K\beta/2} \hspace{4cm}(H>K/4)
\label{quadjfzline}
\end{eqnarray} 
It is evident from the above equation that for $K/6<H<K/4$ the rate of growth of quadrupole fluctuations is progressively slower than that of the correlation length, with the former eventually completely flattening at $K=6H$. For $H<K/6$ the quadrupole fluctuation decays to zero towards zero temperature while the correlation length continues to grow. For the magnetization fluctuations the rate remains the same as above excepting the case $H<K/8$ for which the rate of decay of fluctuations becomes $e^{-2H\,\beta}$. 
\begin{figure}[t!]
\begin{minipage}[b]{0.3\linewidth}
\centering
\includegraphics[width=1.7in,height=1.2in]{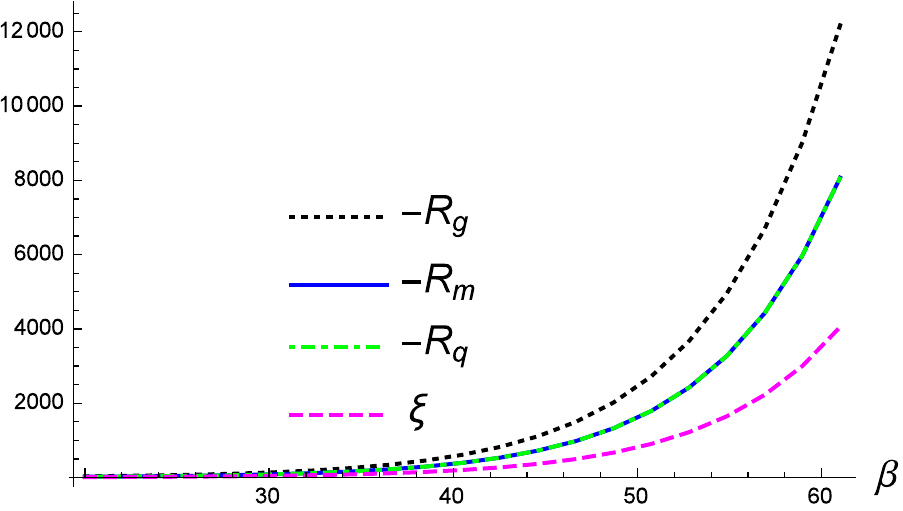}
\end{minipage}
\hspace{0.2cm}
\begin{minipage}[b]{0.3\linewidth}
\centering
\includegraphics[width=1.7in,height=1.2in]{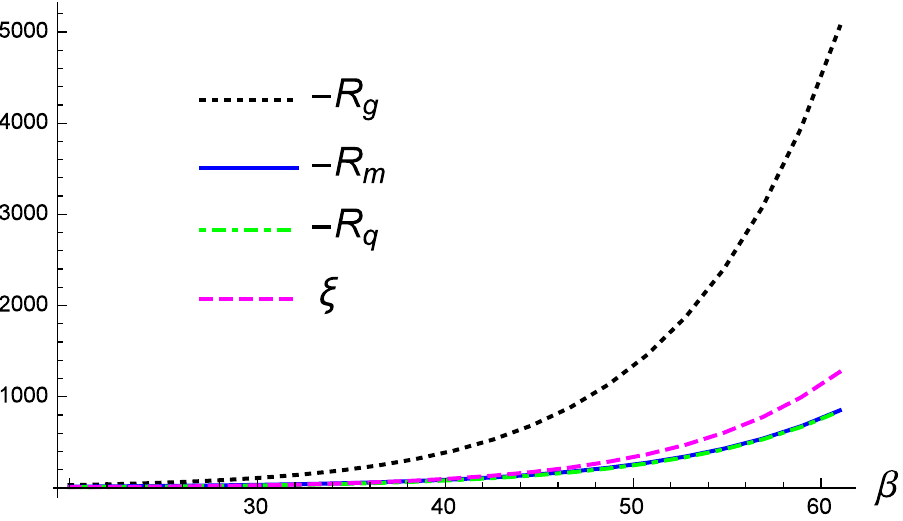}
\end{minipage}
\hspace{0.2cm}
\begin{minipage}[b]{0.3\linewidth}
\centering
\includegraphics[width=1.7in,height=1.2in]{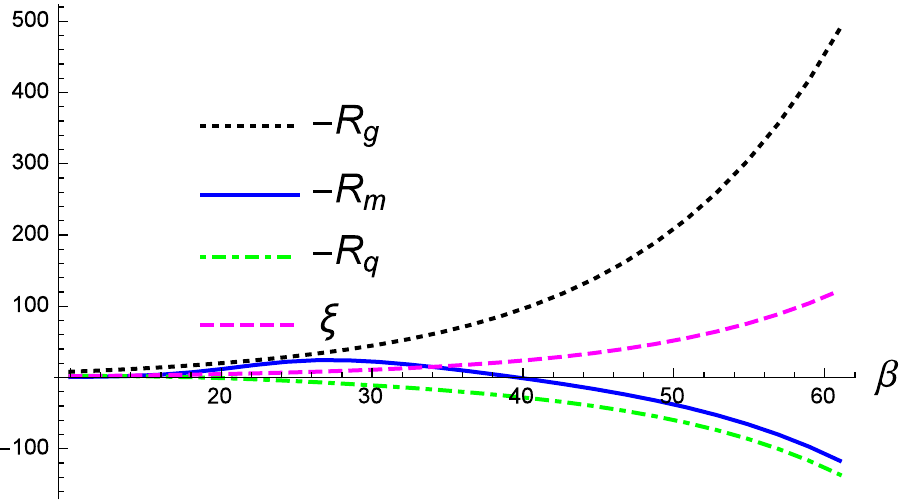}
\end{minipage}
\caption{\small{ Temperature plots of curvatures and correlation length for the pseudocritical $f$ line in the Griffiths model. In $(a)$ $H=0.12,D=0.42,K=0.30$, in $(b)$ $H=0.06,K=0.30,D=0.36$ and in $(c)$ $H=0.04,K=0.30,D=0.34$}}
\label{leop5}
\end{figure}

We now survey the geometry of the Griffiths model, starting with the $f$ line. For $H>K/4$ both $R_m$ and $R_q$ show a negative divergence and overlap with each other at low temperatures. They follow the Ruppeiner equation with $\kappa=1$. On the other hand the three dimensional curvature $R_g$ follows the Ruppeiner equation with $\kappa_3=3/2$. The relation with the correlation length can be worked out from eq.(\ref{freesingfzline}) and is obtained as $R_m,R_q\sim 2\xi$ and $R_g\sim3\xi$. This is shown is fig.\ref{leop5}(a) where $R_m$ and $R_q$ are seen to completely overlap with each other. 
For $K/4<H<K/6$ $R_m$ and $R_q$ still show a negative divergence at the same rate as the correlation length (or $\psi^{-1}$), however the proportionality constant $\kappa$ in the Ruppeiner equation is less than unity and steadily decreases to zero as $H$ approaches $K/6$. Moreover, its strength only depends on the ratio of $K/H$. For the grand curvature $R_g$ the constant $\kappa_3=4$ now. This is shown in fig.\ref{leop5}(b) where the correlation length is now seen to lead the curvatures $R_m$ and $R_q$ which again overlap with each other. We could say that the sectional curvatures $R_m$ and $R_q$ are sensitive to the same statistical interactions which cause the fluctuations to flatten out
as $H$ lowers from $K/4$ to $K/6$. Finally, for $H<K/6$ the quadrupole fluctuations start decaying to zero. At the same time the curvatures $R_m$ and $R_q$ now turn positive and diverge to positive infinity at the same rate as the correlation length. Though the two curvatures follow each other they are not as close in value as the previous cases. The poroportionality constant $\kappa$  now ranges from less than unity to much larger values as $H$ approaches zero while $\kappa_3$ continues to remain $4$. This is shown in fig.\ref{leop5}(c) and is favourably compared with an analogous behaviour of $R_q$ in fig.(\ref{norm6}) in sec.(\ref{hzerojless}). In both cases some of the curvatures show a positive divergence at the same rate as the correlation length. The sign change is concomitant with the decay to zero of quadratic fluctuations. While it is unusual for the state space scalar curvature to undergo a positive divergence at criticality, see \cite{rupprev} for instance, we believe that the same statistics which causes the quadratic fluctuations to decay near criticality also causes a sign change in the curvatures here. We shall analyse this further in a future investigation. 

\begin{figure}[t!]
\begin{minipage}[b]{0.3\linewidth}
\centering
\includegraphics[width=1.7in,height=1.2in]{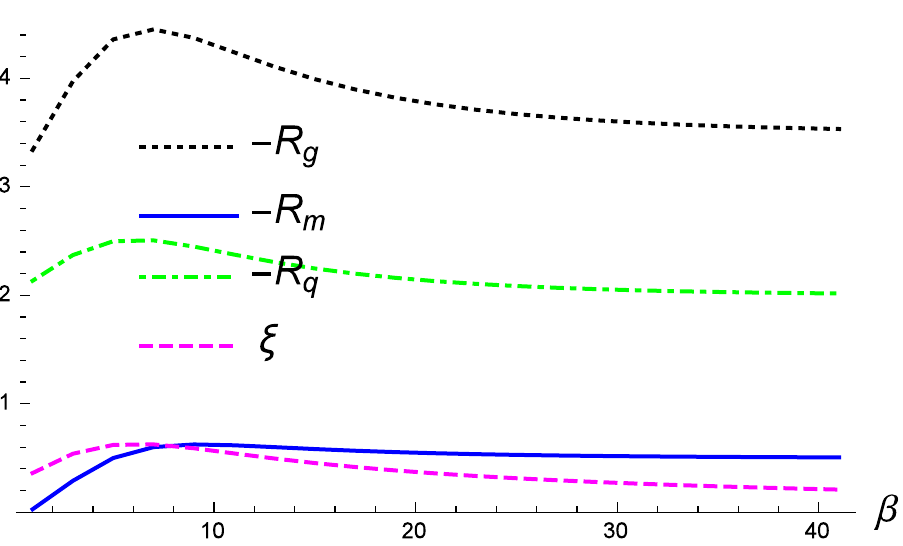}
\end{minipage}
\hspace{0.2cm}
\begin{minipage}[b]{0.3\linewidth}
\centering
\includegraphics[width=1.7in,height=1.2in]{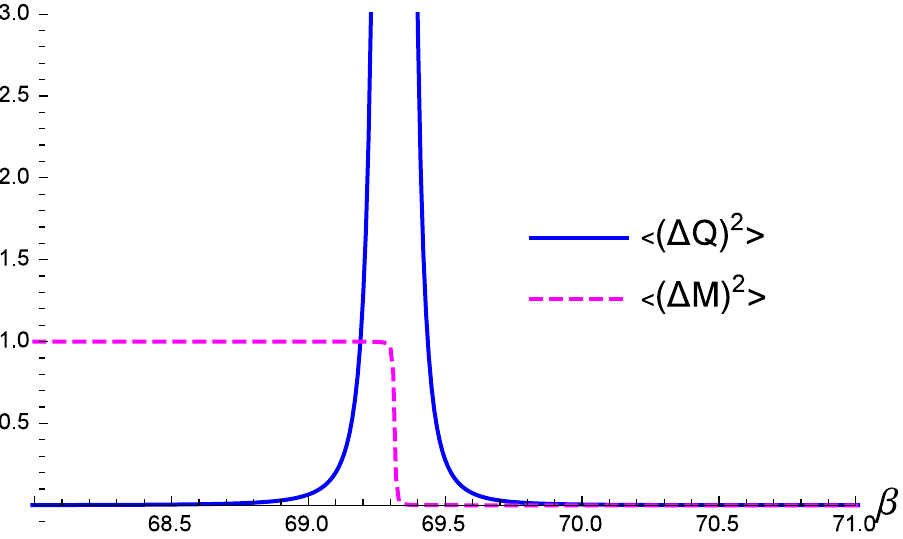}
\end{minipage}
\hspace{0.2cm}
\begin{minipage}[b]{0.3\linewidth}
\centering
\includegraphics[width=1.7in,height=1.2in]{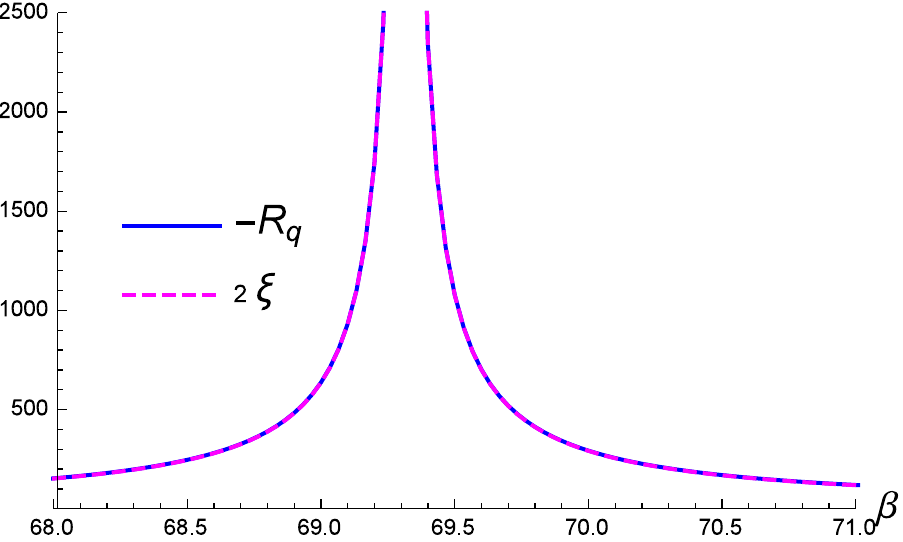}
\end{minipage}
\caption{\small{ $(a)$ Scalar curvatures and correlation length for the Griifiths model with $H=0,K=0.30,D=0.31$. Plots of $(b)$ spin and quadrupole fluctuation moments and $(c)$ $2 \xi$ and $-R_q$ for $H=0.06,K=0.30,D=0.31$}}
\label{leop6}
\end{figure}

The $H=0$ axis does not become critical as mentioned earlier. This is reflected very well by the respective curvatures. Fig.\ref{leop6}(a) is a representative zero field plot with $D<K$. The curvature $R_m$ quickly converges to a value of $1$ while $R_q$ converges to $-2$ and $R_g$ to $-3.5$, with the correlation length decaying to zero. At the triple point $D=K$ the curvatures asymptote to $R_g\to-7.5,R_q\to-4,R_m\to-1$ and the correlation length now does not decay to zero but converges to 1/log($2$). For $D>K$ we report a minor but interesting feature which is ably captured by the geometry of the model. As shown in fig.\ref{leop6}(b) the quadratic fluctuation after initially decaying to zero undergoes a spike in fluctuation. Exactly in the middle of this spike the magnetization fluctuation sharply drops from $1$ to $0$. The curvature $R_q$ remains exactly twice the correlation length as can be seen from fig.\ref{leop6}(c). The curvature $R_m$ and $R_g$ on the other hand jump to large positive values at the temperature at which the magnetization fluctuation drops to near zero values (not shown in the figure).
\begin{figure}[h!]
\begin{minipage}[b]{0.3\linewidth}
\centering
\includegraphics[width=1.7in,height=1.2in]{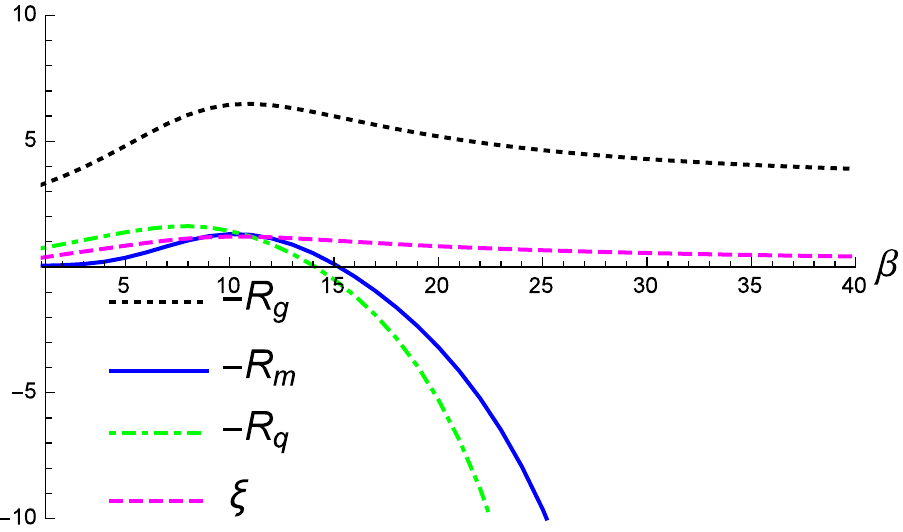}
\end{minipage}
\hspace{0.2cm}
\begin{minipage}[b]{0.3\linewidth}
\centering
\includegraphics[width=1.7in,height=1.2in]{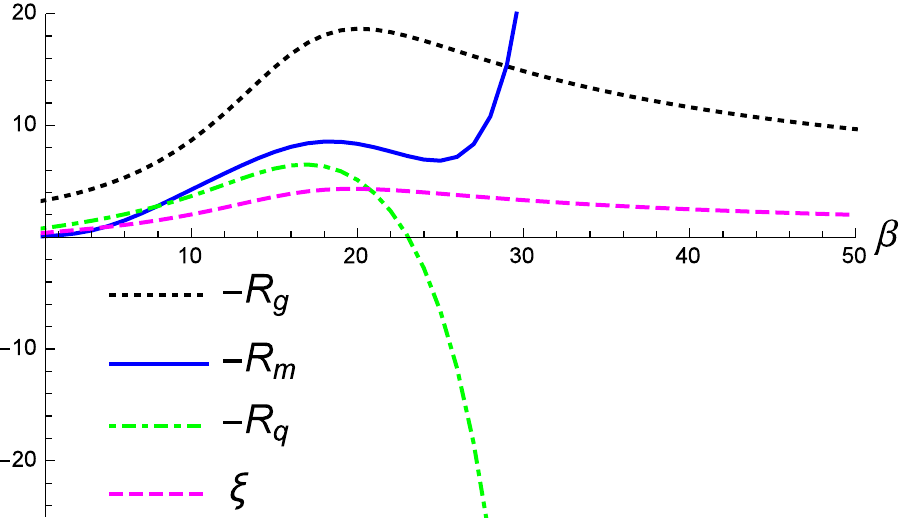}
\end{minipage}
\hspace{0.2cm}
\begin{minipage}[b]{0.3\linewidth}
\centering
\includegraphics[width=1.7in,height=1.2in]{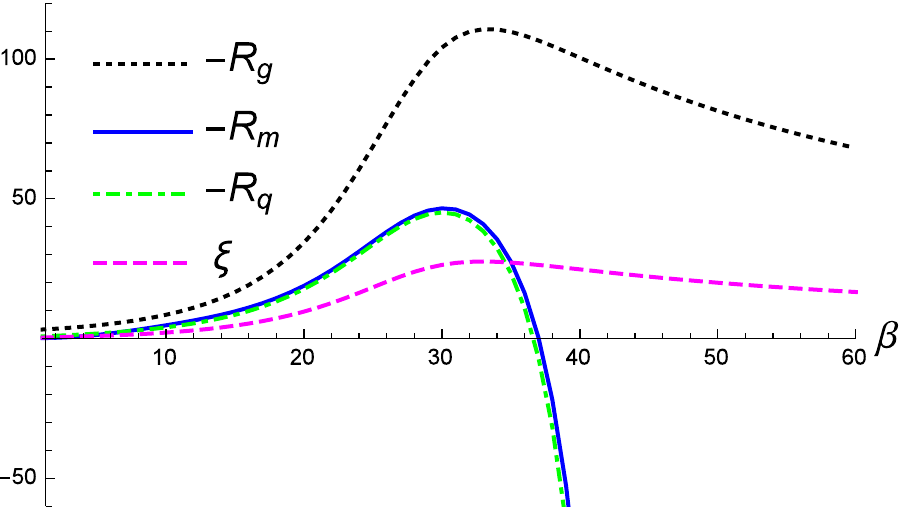}
\end{minipage}
\caption{\small{Temperature plots of scalar curvatures and the correlation length for the Griffiths model. In all sub-figures, $H=0.12,K=0.30$ and $D$ values are less than but progressively closer to the $f$-line at $D=K+H$. In $(a)$ $D=0.36$, in $(b)$  $D=0.41$ and in $(c)$ we have $D=0.419$.}}
\label{leop7}
\end{figure}
   
 We turn now to non-zero $H$ values in regions away from the $f$ line. In fig.(\ref{leop7}) we take a representative selection of plots for a fixed value of $H$ and different values of $D$ upto the $f$ line and to its left. In all the sub-figures the curvatures $R_m$ and $R_q$ show differing behaviour perhaps signalling the different nature of underlying statistics, a question we shall be addressing in a future investigation. On the other hand the grand curvature $R_g$ appears to always correspond well with the correlation length, albeit at an elevated value. To the right of the $f$ line all the curvatures change sign and diverge to positive infinity (not shown).
\section{Conclusions}
\label{conclu}
 In this work we have undertaken an extensive analysis of the thermodynamic geometry associated with the one dimensional BEG model and its limiting cases. The BEG model which is extensively used to model interacting systems with two competing ordering processes  preserves much of the richness of its phase structure in the one dimensional case. In addition, being exactly solvable in one dimension it offers an excellent opportunity to comprehensively probe its geometry. 
 
 The three dimensional state space of the BEG model has been systematically sectioned into two co-dimension one  hypersurfaces, namely  constant $H$-surface and the constant $D$-surface. The associated scalar curvatures $R_q$ and $R_m$ are found to be relevant to the fluctuations in the quadrupolar order parameter $Q$ and the magnetic moment $M$ respectively. For $H=0$ the spin and quadrupolar order parameters have separate correlation lengths $\xi_1$ and $\xi_2$ and, remarkably enough, it is seen that the two sectional curvatures encode these separately. The curvatures $R_q$ and $R_m$ not only follow the Ruppeiner equation near (pseudo)criticality with $\kappa=1$ they also satisfy the weak conjecture of Ruppeiner showing a very good correspondence with the correlation length away from criticality and even in non-critical situations. Making use of the Ruppeiner equation we are also able to ascertain the  scaling form of the free energy of the BEG model near criticality and pseudocriticality. The three dimensional scalar curvature $R_g$ also efficiently encodes the interactions in the system, especially in the limit of the Griffiths model where the bilinear spin coupling is set to zero. One of the key messages in our work is that for higher dimensional parameter spaces the relevant sectional curvatures contain essential information not found in the full scalar curvature. Therefore, for higher dimensional parameter spaces, in addition to the full scalar curvature the relevant hypersurface geometries must also be explored to gain a fuller picture of underlying statistical interactions.

 While $R_g$ is proportional to the inverse of singular free energy near criticality, the proportionality constant $\kappa_3$, though of order unity, is seen to vary depending on the parameter range. This variation in the proportionality constant does not seem to accord with its universal constant value as worked out in \cite{rupphigher} and also verified there for the mean field case. We also extensively record the sign changes in the scalar curvatures. While in some cases the sign changes are found to encode the peculiar nature of fluctuations in other cases we are not able arrive at a more fundamental understanding of the same. We hope to address these unresolved issues in a future investigation.

\section{Acknowledgements}

We thank George Ruppeiner for fruitful and encouraging discussions during the early stages of the work. AS thanks Ritu Sharma and Rishabh Jha for discussions and DST, Govt. of India for support through grant no. MTR/2017/001001.

\begin{appendices}

\begin{eqnarray}
\mathcal{N}_1 &=& -2 e^{-J\,\beta} \,(37 e^{2 \beta  J}+4 e^{16 \beta  J}+5 W e^{\beta  J}+(6 W+24) e^{15 \beta  J}+(24 W+334) e^{14 \beta  J}+\nonumber\\ &&(37 W+74) e^{3 \beta  J}+(54 W+106) e^{4 \beta  J}+(106 W+464) e^{5 \beta  J}+(274 W+980) e^{13 \beta  J}+\nonumber\\&&(304 W+1100) e^{7 \beta  J}+(316 W+480) e^{6 \beta  J}+(396 W+1702) e^{12 \beta  J}+\nonumber\\ &&(676 W+1671) e^{8 \beta  J}+(703 W+1684) e^{9 \beta  J}+(733 W+1868) e^{11 \beta  J}+\nonumber\\&&(740 W+2589) e^{10 \beta  J}+5)\nonumber\\
\mathcal{D}_1 &=& W (6 e^{2 \beta  J}+2 e^{7 \beta  J}+2 W e^{\beta  J}+W e^{6 \beta  J}+(4 W+27) e^{5 \beta  J}+\nonumber\\&&(5 W+4) e^{4 \beta  J}+(6 W+13) e^{3 \beta  J}+2)^2\nonumber\\
W&=&\sqrt{e^{-2 \beta  J}+8 e^{\beta  J}}
\label{A1}
\end{eqnarray}

\begin{eqnarray}
\mathcal{N}_2 &=& \mathcal{A}\,\mathcal{B}\nonumber\\
\mathcal{A}&=& -2 e^{-\beta  J} \left(e^{3 \beta  D}+(W+6) e^{\beta  (D+2 J)}+3 (W+2) e^{\beta  D}+\right.\nonumber\\&&\left.(W+6) e^{2 \beta  D}+e^{6 \beta  J}+(W+3) e^{4 \beta  J}+(2 W+3) e^{2 \beta  J}+W+1\right)\nonumber\\
\mathcal{B}&=&6 e^{\beta  (5 D+J)}+e^{\beta  (D+9 J)}+4 e^{4 \beta  D+3 \beta  J}+e^{3 \beta  D+5 \beta  J}+3 e^{2 \beta  D+7 \beta  J}+\nonumber\\&&14 (W+2) e^{\beta  (D+3 J)}+ 10 (W+3) e^{\beta  (D+5 J)}+5 (W+5) e^{2 \beta  D+5 \beta  J}+2 (W+6) e^{\beta  (D+7 J)}+\nonumber\\&&(6 W+9) e^{\beta  (D+J)}+(6 W+28) e^{\beta  (4 D+J)}+(6 W+34) e^{3 \beta  (D+J)}+(13 W+27) e^{\beta  (2 D+J)}+\nonumber\\&&(14 W+41) e^{\beta  (3 D+J)}+(18 W+49) e^{2 \beta  D+3 \beta  J}+e^{11 \beta  J}+(W+1) e^{\beta  J}+\nonumber\\&&(W+5) e^{9 \beta  J}+2 (2 W+5) e^{7 \beta  J}+2 (3 W+5) e^{5 \beta  J}+(4 W+5) e^{3 \beta  J}\nonumber\\
\mathcal{D}_2&=&W \left(e^{\beta  D}+e^{2 \beta  J}+W+1\right) \left(2 e^{3 \beta  D}+e^{2 \beta  (D+J)}+(W+6) e^{\beta  (D+2 J)}+3 (W+2) e^{\beta  D}+\right.\nonumber\\&&\left.(2 W+7) e^{2 \beta  D}+e^{6 \beta  J}+(W+3) e^{4 \beta  J}+(2 W+3) e^{2 \beta  J}+W+1\right)^2\nonumber\\
W&=&\sqrt{6 e^{\beta  D}+e^{2 \beta  D}-2 e^{\beta  (D+2 J)}+2 e^{2 \beta  J}+e^{4 \beta  J}+1}
\label{A2}
\end{eqnarray}

\end{appendices}

\end{document}